\documentclass[prd]{revtex4}
\usepackage{color,graphicx}
\usepackage{epsf}

\usepackage{amsbsy}
\usepackage{amssymb}
\usepackage{amsmath}
\begin{document}

\title{Tkachenko modes in rotating neutron stars:\\ the effect of compressibility and implications for pulsar timing noise}

\author{B. Haskell$^{1,2}$}
\address{$^1$School of Mathematics, University of Southampton, Southampton, SO17 1BJ, UK}
\address{$^2$ Astronomical Institute ``Anton Pannekoek'', University of Amsterdam, Science Park 904, 1098 XH Amsterdam, Netherlands}

\date{\today}

\voffset 0.5 truein

\def\be{\begin{equation}}
\def\ee{\end{equation}}
\def\bea{\begin{eqnarray}}
\def\eea{\end{eqnarray}}
\def\bear{\begin{eqnarray}}
\def\eear{\end{eqnarray}}
\def\beq{\begin{eqnarray}}
\def\eeq{\end{eqnarray}}
\def\v{{\rm v}}
\def\pn{{\rm pin}}
\def\n{{\rm n}}
\def\x{{\rm x}}
\def\p{{\rm p}}
\def\s{{\rm s}}
\def\N{{\rm N}}
\def\S{{\rm S}}
\def\mun{{\mu_\n}}
\def\mus{{\mu_\s}}

\begin{abstract}
Long wavelength oscillations (Tkachenko waves) of the triangular lattice of quantized vortices in superfluid neutron stars  have been suggested as one of the possible explanations for the timing noise observed in many radio pulsars, in particular for the 100-1000 day variations in the spin of PSR 1828-11.  Most studies to date have, however, been based on the hydrodynamics developed for superfluid Helium. In this paper we extend the formulation to a two fluid neutron and proton system, relevant for neutron star interiors and include the effect of chemical coupling, compressibility and mutual friction between the components. In particular we find that chemical coupling and compressibility  can have a drastic effect on the mode structure. However, for the slower pulsars rotating at 1-10 Hz (such as PSR B1828-11), most choices of parameters in the equation of state lead to Tkachenko oscillations with frequencies in the correct range to explain the timing noise. We also investigate the case of more rapidly rotating pulsars (above 100 Hz) for which we find that there is a vast portion of parameter space in which there are no Tkachenko modes, but only modified sound waves at much higher frequencies.

\end{abstract}

\maketitle

\section{Introduction}

A growing number of radio pulsars have now been observed for over a decade (some for more than 30 years) and are, in general, very stable rotators. However, many pulsars also exhibit timing irregularities, such as ``glitches'', which are sudden increases in the rotation rate, and ``timing noise'', a general term which refers to low frequency quasi-periodic structures that appear in the timing residuals, once the ``regularly'' pulsating solution has been removed.  In particular, while the irregularities in younger pulsars are dominated by the recovery from glitch events, for a handful of older pulsars there is growing evidence for long-period ($\sim$ 100-1000 days) oscillations in the timing residuals \cite{pulsars}.  In some cases these periodicities, and the correlated pulse shape changes, can be partially explained by neutron star free-precession, and one of the best examples of this is PSR B1828-11 which shows significant periodicity at $\sim 256$ days and $\sim 511$ days \cite{stairs}.  
However there are theoretical arguments stating that mutual friction between the interior superfluid components of the star would damp out any precessional motion on a short timescale \cite{Shaham,Sed1} (although Glampedakis et al. \cite{Kostas1} have shown that short wave length instabilities in the pinned superfluid could cast an element of doubt on such conclusions).  Furthermore recent work shows that some pulsars may be switching abruptly between two different states with different spin-down rates, thus giving rise to the observed timing behaviour \cite{Kram}.

Noronha and Sedrakian \cite{Sed2}, following earlier suggestions by Ruderman \cite{Rud}, indicated that an alternative explanation for the observed long term periodicity could be the propagation of Tkachenko waves in the star. In fact, it has been suggested that Tkachenko waves excited by glitches may be driving precession in one of the X-ray Dim Isolated Neutron stars (XDINs), RX J0720.4-3125 \cite{Popov}.

Neutron star interiors are expected to contain charge neutral superfluids that rotate by forming an array of quantized vortices. In their lowest energy state the vortices form a two-dimensional triangular lattice that can support elastic oscillations, Tkachenko waves \cite{Tka}, that have been studied extensively, both theoretically and  experimentally, in superfluid $^4$He (see e.g. \cite{Andereck} for a review) and recently in Bose-Einstein condensates (BECs) (see e.g. \cite{Baym1} and \cite{Baym2}). The undamped propagation of Tkachenko waves in a neutron star would lead to periodic variations in the angular momentum of the superfluid which, due to coupling to the crust, would lead to variations in the observed rotation rate.
In order to ascertain if this is a viable hypothesis it is crucial to understand how the detailed microphysics of neutron star interiors affects the propagation of the modes. Most studies to date have been based on the hydrodynamical theory of Tkachenko waves developed by Baym and Chandler \cite{CB1,CB2} for superfluid $^4$He. In this case the fluids can be treated, to a good degree of approximation, as incompressible, given that the rotation rate is always well below the sound wave frequency (note, however, that in BECs, compressibility has a strong effect, due to the interactions being much weaker and the sound speed much lower than in Helium), and the system can be described as a condensate (the "superfluid") coupled to a "normal" fluid which consists, loosely speaking, of the thermal excitations of the system. 

In a realistic neutron star, on the other hand, one must take into account not only the effects of rapid rotation, but also the presence of several massive fluids, describing the flow of electrons, protons, superfluid neutrons (and their excitations at finite temperature) and possibly exotic particles such as hyperons or deconfined quarks, which cannot, in general, be assumed to be incompressible. Furthermore, one must consider various dissipative processes that damp out the oscillations. It is well known that in a multifluid system there will, in general, be many more dissipation channels than in a simple one-fluid flow described by the Navier-Stokes equation \cite{EP,Monster}. 

Solving the full problem is clearly a daunting task, so in this paper we shall make a series of simplifying assumptions. We take a system of two fluids: the superfluid neutrons and a charged component of protons and electrons, which are locked by the Coulomb interaction on a much shorter timescale than the dynamical timescales considered here. This assumption is justified as the frequencies of the modes we calculate (Tkachenko waves, sound waves and inertial waves) are always significantly lower than the electron-proton cyclotron and plasma frequencies, i.e. lower than $\approx 10^{15}$ Hz (\cite{Mendell}). Both fluids are assumed to be compressible, and we shall assume some simplified analytic models for the equation of state, derived from Haskell et al. \cite{rmode}. Furthermore we will only consider the damping due to superfluid mutual friction, which has been found to have a significant effect on the mode propagation \cite{Sed1}. 

The paper is structured as follows. In Section II we present the formalism for studying the oscillations of a two-fluid neutron star with mutual friction and vortex lattice elasticity. In Section III we perform a plane wave analysis of the oscillation spectrum in the incompressible case and in Section IV we present the more realistic compressible case. As we shall see compressibility and chemical coupling between the components can profoundly alter the nature of the Tkachenko waves, leading in some cases to much shorter periods of oscillation, close to the rotation period of the star, which are not consistent with the observed periodicities in pulsar timing residuals. Finally in Section V we outline our conclusions.

\section{Two fluid equations of motion}

Our starting points will be the multi-fluid formulation of superfluid hydrodynamics of Andersson and Comer \cite{Monster}, and the Baym-Chandler formalism for including the effects of vortex lattice elasticity  in the study of superfluid $^4$He \cite{CB1}. Let us consider a two fluid system of neutrons and protons (which we assume locked to the electrons \cite{Mendell}) and write the Euler equations for the neutrons, in a frame rotating with the star at fixed angular velocity $\Omega$ and in the absence of external forces and mutual friction. Following \cite{Monster} this takes the form:
\be
(\partial_t + v^\n_j\nabla^j)(v^\n_i+\varepsilon_\n w^{\p\n}_i)+2\epsilon_{ijk}\Omega^j v_\n^k + \nabla_i (\tilde{\mu}_\n+\phi)+\varepsilon_\n w^j_{\p\n}\nabla_i v^\n_j =0\label{eulern}
\ee
Where $v^\n_i$ is the neutron velocity, $w_i^{\p\n}=v_i^\p-v_i^\n$, with $v_i^\p$ the proton velocity, $\varepsilon_\n$ is the entrainment parameter for the neutrons \cite{Prix}, $\tilde{\mu}_\n$ is the chemical potential per unit mass of the neutrons and $\phi$ is the gravitational potential. Note that we assume summation of repeated indices and assume the neutron and proton masses equal, $m_\n=m_\p=m$.
In the above equation we have not yet imposed that the neutrons be superfluid. To do this we must require that the fluid rotates by forming an array of singly quantized vortices, and that averaging over a large number of such vortices gives rise to the macroscopic vorticity $\omega^i$ of the fluid  via the relation:
\be
\omega^i = \kappa n_\v \hat{\kappa}^i =\frac{1}{m} \epsilon^{ijk}\nabla_j (v^\n_k+\varepsilon^\n w^{\p\n}_k)
\label{circolo}
\ee
where $\hat{\kappa}^i$ is a unit vector along the direction of the vortex array, $\kappa=h/2 m_\n=1.99 \times 10^{-3}$ cm$^2$ s$^{-1}$ is the quantum of circulation and $n_\v$ is the number of vortices threading a unit surface. It is important to remark here that the quantization condition on the circulation is a condition on the momentum of the neutron fluid $p_i^\n= m (v^\n_i+\varepsilon^\n w^{\p\n}_i)$, and not on its velocity $v_i^\n$ (which will not in general be aligned with $p_i^\n$ due to the entrainment).
From equation (\ref{circolo}) one can derive the equation of motion for the circulation
\be
\partial_t \omega_i+\epsilon_{ijk}\epsilon^{klm}\nabla^j \omega_l v^\v_m=0\label{circolo1}
\ee
and a conservation equation for the vortex number
\be
\partial_t n_\v +\nabla_i (n_\v v^i_\v) =0\label{circolo2}
\ee
with $v^i_\v$ the macroscopically averaged vortex velocity. 
It is possible to show (\cite{supercon}) that, in order for equations (\ref{circolo1}) and (\ref{circolo2}) to be staisfied it is necessary to add a "Magnus force" term to the right hand side of equation (\ref{eulern}), which thus takes the form
\be
(\partial_t + v^\n_j\nabla^j)(v^\n_i+\varepsilon_\n w^{\p\n}_i)+2\epsilon_{ijk}\Omega^j v_\n^k + \nabla_i (\tilde{\mu}_\n+\phi)+\varepsilon_n w^j_{\p\n}\nabla_i v^\n_j =\kappa n_\v \epsilon_{ijk} \hat{\kappa}^j (v_\n^k-v_\v^k)\label{eulern2}
\ee
It is clear from equation (\ref{eulern2}) that in the absence of other forces the vortices will be forced to move with the superfluid neutron condensate.
The presence of vortices will, however, also affect the proton fluid, which will experience a drag force of the form $\rho_\n\kappa n_\v \mathcal{R} (v^i_\v-v^i_\p)$, where the exact nature of the process giving rise to the drag is encoded in the dimensionelss parameter $\mathcal{R}$. In a neutron star there are, in fact, a variety of mechanisms that can produce a dissipative drag: scattering of electrons off vortex cores is likely to be the dominant process in the core \cite{HV,Trev}, while in the crust the main contribution is due to interactions with the lattice phonons \cite{Jones1} and the excitation of vortex Kelvin waves \cite{EB, Jones2}. The diverse nature of these processes leads to the drag parameter spanning several orders of magnitude in the different regions of a neutron star interior (from as low as $\mathcal R\approx 10^{-10}$ to  $\mathcal R\approx 1$). We shall thus treat $\mathcal R$ as a free parameter and investigate how its variations affect the modes.

The Euler equations for the proton fluid take the form:
\be
(\partial_t + v^\p_j\nabla^j)(v^\p_i-\varepsilon_\p w^{\p\n}_i)+2\epsilon_{ijk}\Omega^j v_\p^k + \nabla_i (\tilde{\mu}_\p+\phi)+\varepsilon_\p w^j_{\n\p}\nabla_i v^\p_j =\kappa n_\v \frac{(1-x_\p)}{x_\p}\mathcal{R} (v^\v_i-v^\p_i)\label{eulerp}
\ee
where $x_\p=\rho_\p/(\rho_\n+\rho_\p)$. $\tilde{\mu}_\p$ and $\varepsilon_\p$ are now the chemical potential per unit mass and entrainment parameter of the protons, such that $\varepsilon_\p=\varepsilon_\n (1-x_\p)/x_\p$ \cite{Prix}.
We also need an equation of motion for the vortex lines which, if we assume that they have negligible inertia, takes the form of a force balance between the Magnus force, the drag force and the elastic force exerted by the lattice \cite{CB1}:
\be
\rho_\n\kappa n_\v \epsilon_{ijk}\hat{\kappa}^j (v_\v^k-v_\n^k)+\rho_\n\kappa n_\v \mathcal{R}(v^\p_i-v^\v_i)-\rho_\n\sigma_i=0\label{forcev}
\ee 
where $\sigma_i$ represents the contribution due to lattice elasticity and takes the form:
\be
\sigma_i=\frac{\mu_\v}{\rho_\n}\left[2\nabla^\perp_i (\nabla_\perp^j \epsilon_j)-(\nabla^2_\perp)\epsilon_i\right]
\ee
where $\epsilon_i$ is the displacement of the vortex line from its equilibrium position, $\nabla_\perp^j$ is the gradient perpendicular to the direction of the array and $\mu_\v=\rho_\n\kappa^2 n_\v / 16 \pi$ is the shear modulus of a triangular vortex lattice \cite{Tka}. Note that the above expression only describes the linear order corrections in the lattice displacements, which are assumed to be small. Furthermore we are neglecting the contribution of vortex bending, which would give rise to Kelvin waves propagating along the vortex lines. Note that this could be accounted for by including a vortex "tension" term in $\sigma_i$, which we denote $\sigma^T_i$, of the form:
\be
\sigma^{T}_i=-\frac{\rho_\n \kappa^2n_\v}{8\pi}\ln\left(\frac{b}{a}\right)\frac{\partial^2\epsilon_i}{\partial z^2}
\ee
where $a$ is the vortex core radius, $b$ is the inter-vortex spacing for a triangular lattice and the $z$ axis is taken along the rotation axis of the star.

The continuity equations for neutrons and protons take the form
\beq
&&\partial_t \rho_\n+\nabla^i(\rho_\n v^\n_i)=0\\
&&\partial_t \rho_\p+\nabla^i(\rho_\p v^\p_i)=0
\eeq
and the gravitational potential obeys the Poisson equation
\be
\nabla^2 \phi=4\pi G (\rho_\n+\rho_\p)
\ee
where $G$ is the gravitational constant. Finally to solve the problem we need to supply an equation of state for the system. As we shall examine different cases we delay the discussion of the equation of state to the following sections and move on to discussing perturbations of the multi-fluid equations of motion presented above.

\subsection{Perturbations}

In order to keep the problem tractable we shall consider linear perturbations of a background in which the two fluids rotate together with uniform angular velocity $\Omega$. For such a background equation (\ref{circolo}) takes the form:
\be
\kappa n_\v=2\Omega
\ee
and the perturbed Euler equations can be written, in a frame co-rotating with the star, as:
\beq
&&\partial_t (\delta v_i^\n + \varepsilon_\n \delta w_i^{\p\n}) + 2\epsilon_{ijk} \Omega^j \delta v_\n^k +\nabla_i \delta \tilde{ \mu}_\n = -2\Omega\mathcal{R} (\delta v^\v_i-\delta v^\p_i)-\sigma_i\\
&&\partial_t (\delta v_i^\p - \varepsilon_\p \delta w_i^{\p\n}) + 2\epsilon_{ijk} \Omega^j \delta v_\p^k +\nabla_i \delta \tilde{ \mu}_\p = 2\Omega \frac{(1-x_\p)}{x_\p}\mathcal{R} (\delta v^\v_i-\delta v^\p_i)\label{eulers}
\eeq
where we have made the Cowling approximation, i.e. neglected the perturbations of the gravitational potential $\delta\phi$. Note that as a consequence of the extra elastic term in the force balance equation for the vortices (\ref{balo}), the forces on the right hand side of the Euler equations are no longer symmetric and the vortex elasticity term only acts on the neutron superfluid. The elastic force $\sigma_i$ can be written as
\be
\sigma_i=c_T^2\left[2\nabla^\perp_i (\nabla_\perp^j \epsilon_j)-(\nabla^2_\perp)\epsilon_i\right]
\ee
where we have defined the Tkachenko wave speed $c_T^2=\kappa \Omega/8\pi$ and are assuming the vortices to be in equilibrium in the background, such that $\epsilon_j^{\mathrm{BKG}}=0$. We assume all vortex displacements to be perturbed quantities, and write $\epsilon_i$ in place of $\delta\epsilon_i$, unless otherwise specified, and thus consider $\sigma_i$ to also be a perturbed quantity. As we are dealing with linearised elasticity and the displacement vectors $\epsilon_i$ it would be natural to consider Lagrangian perturbations of the two-fluid equations of motion, given that in general one would have that $\Delta v_\v^i=\partial_t \epsilon_\v^i$. However, given that we are working in a rotating frame, and have assumed that the fluids (and thus the vortices) are moving together in the background, one has that $\Delta v_\v^i=\delta v_\v^i$. We can thus continue to work with Eulerian perturbations, which simplifies somewhat the problem.

The equation of force balance for the vortices (\ref{forcev}) can be cast in the form:
\be
\delta v_i^\v=\delta v_i^\p+\frac{\mathcal{R}}{1+\mathcal{R}^2}\epsilon_{ijk}\hat{\kappa}^j \delta w^k_{\p\n}-\frac{\mathcal{R}}{2\Omega(1+\mathcal{R}^2)}\sigma_i+\frac{\hat{\kappa}_i}{1+\mathcal{R}^2}(\delta w_j^{\p\n}\hat{\kappa}^j)-\frac{\delta w_i^{\p\n}}{1+\mathcal{R}^2}-\frac{\epsilon_{ijk}\hat{k}^j\sigma^k}{2\Omega(1+\mathcal{R}^2)}
\label{balo}\ee
and the perturbed continuity equations, in the absence of reactions, take the form
\beq
&&\partial_t \delta \rho_\n+\nabla^i(\rho_\n \delta v^\n_i)=0\\
&&\partial_t \delta \rho_\p+\nabla^i(\rho_\n \delta v^\p_i)=0\label{continuo}
\eeq
Following \cite{fmode} we can combine equations (\ref{eulers}) to obtain an Euler equation for the "total" velocity $v_i=(1-x_\p)v^\n_i+x_\p v^\p_i$:
\be
\partial_t \delta v_i +\frac{1}{\rho}\nabla_i\delta p-\frac{\delta\rho}{\rho}\nabla_i p+2\epsilon_{ijk}\Omega^j\delta v^k=-(1-x_\p)\sigma_i
\label{tutto1}\ee
and one for $w_i^{\p\n}$:
\be
(1-\bar{\varepsilon})\partial_t w_i^{\p\n}+\nabla_i\delta\beta=-2\Omega\tilde{\mathcal{B}}^{'}\epsilon_{ijk}\hat{\kappa}^j\delta w_{\p\n}^k+2\Omega\tilde{\mathcal{B}}\epsilon_{ijk}\hat{\kappa}^j\epsilon^{klm}\hat{\kappa}_l\delta w_m^{\p\n}+\sigma_i
\label{tutto2}\ee
where we have defined the total pressure, such that $\nabla_i P=\rho_\n\nabla_i\tilde{\mu}_\n+\rho_\p\nabla_i\tilde{\mu}_\p$, the entrainment parameter $\bar{\varepsilon}=\varepsilon_\p+\varepsilon_\n$, $\delta\beta=\delta\tilde{\mu}_\p-\delta\tilde{\mu}_\n$ and the mutual friction parameters $\tilde{\mathcal{B}}^{'}=1-\mathcal{R}^2/[x_\p(1+ \mathcal{R}^2)]$ and $\tilde{\mathcal{B}}=\mathcal{R}/[x_\p(1+ \mathcal{R}^2)]$.
The perturbed continuity equations (\ref{continuo}) can be cast in the form:
\beq
&&\partial_t \delta \rho+\nabla_i (\rho \delta v^i)=0\\
&&\partial_t\delta x_\p+\frac{1}{\rho}\nabla_j[x_\p(1-x_\p)\rho\delta w^j]+\delta v^j \nabla_j x_\p=0
\label{tutto3}
\eeq
As we shall see in the following, this formulation can be advantageous when discussing the compressible problem.

\section{ The incompressible case}

In order to make contact with previous results, let us consider first of all the case of incompressible fluids, such that $\delta\rho=0$ and the continuity equations reduce to
\be
\nabla^i \delta v_i^\p=\nabla^i \delta v_i^\n=0
\ee
We consider plane waves, such that a perturbed quantity $\delta f_i(\mathbf{x},t)$ takes the form $\delta f_i(\mathbf{x},t)=\bar{f}_i\exp(ik_ix^i-i\omega t)$, with $\bar{f}_i$ a constant amplitude. Without loss of generality we choose our coordinate system such that the $z$ axis points along the rotation axis and such that the wave vector $k^i$ lies in the $x-z$ plane, i.e. $\mathbf{k}=(k\sin\theta,0,k\cos\theta)$. The equations of motion can thus be written as:
\beq
&&-i\omega \bar{v}_i^\n(1-\varepsilon_\n) -i\omega \varepsilon_\n \bar{v}_i^{\p} + 2\epsilon_{ijk} \Omega^j \bar{v}_\n^k +i k_i  \bar{ \mu}_\n = 2\Omega\mathcal{R} (i\omega\bar{\epsilon}_i+\bar{v}^\p_i)-\tilde{\sigma}_i\label{eulersk1}\\
&& -i\omega \bar{v}_i^\p(1-\varepsilon_\p) -i\omega \varepsilon_\p \bar{v}_i^{\n} + 2\epsilon_{ijk} \Omega^j \bar{v}_\p^k +i k_i \bar{ \mu}_\p = -2\Omega \frac{(1-x_\p)}{x_\p}\mathcal{R} (i\omega\bar{\epsilon}_i+\bar{v}^\p_i)\\
&&-i\omega\bar{\epsilon}_i-\bar{v}_i^{\p}-\mathcal{B}\epsilon_{ijk}\hat{\kappa}^jw^k_{\p\n}+\mathcal{B}\tilde{\sigma}_i-\frac{\hat{\kappa}_i}{(1+\mathcal{R}^2)}w_j^{\p\n}\hat{\kappa}^j+\frac{w_i^{\p\n}}{(1+\mathcal{R}^2)}+\frac{\epsilon_{ijk}\hat{\kappa}^j\tilde{\sigma}^k}{(1+\mathcal{R}^2)}=0\label{forceb}\\
&&k_j v^j_\p=k_j v^j_\n=0
\label{eulersk}
\eeq
where we have defined $\mathcal{B}=\mathcal{R}/(1+\mathcal{R}^2)$, $\tilde{\sigma}_i=\sigma_i/2\Omega$ and to simplify notation we have defined $\bar{\mu}_\x$ as the amplitude of $\delta\tilde{\mu}_\x$, with $\x=\n,\p$. Finally the vortex elasticity contribution takes the form:
\be
\boldsymbol{\sigma}=\mathbf{q}\cdot \boldsymbol{\epsilon}\;\;\;\;\mbox{with}\;\;\;\;\boldsymbol{q}=(-(c_T k\sin\theta)^2,(c_T k\sin\theta)^2,0) 
\ee
In order to obtain the dispersion relation for the modes of the system we thus need to solve the characteristic equation det$|K_{ij}|$=0, where $K_{ij}$ follows from equations (\ref{eulersk1}-\ref{eulersk}) and is given in equation (\ref{kappa1}).

In the undamped case, neglecting the effect of entrainment ($\varepsilon_\n=\varepsilon_\p=0$) one obtains, as expected, two families of modes, the inertial modes
\be
\omega^2=4\Omega^2 (\cos\theta)^2
\ee
and the Tkachenko waves
\be
\omega^2=4\Omega^2(\cos\theta)^2+c_T^2 k^2(\sin\theta)^4-\frac{1}{4}\frac{c_T^4 k^4}{\Omega^2}(\sin\theta)^4\approx4\Omega^2|\cos\theta|^2+c_T^2 k^2(\sin\theta)^4
\label{mod1}\ee
where we are assuming that $c_T^2k^2<<\Omega^2$. This will always be the case if we consider typical pulsar spin rates from a few Hz to a few hundred Hz and long wavelength oscillations across the whole superfluid region, such that $k\approx 10^{-5}\sim 10^{-6}$ cm$^{-1}$.
For propagation perpendicular to the rotation axis ($\cos\theta=0$) one then obtains the well known Tkachenko wave dispersion relation
\be
\omega=\pm c_T k
\label{mod2}
\ee

\subsection{The effect of entrainment}

Let us now still consider undamped propagation of the modes, but include the effect of entrainment. Clearly introducing coupling between the two fluids profoundly alters the nature of the modes and leads, in the $c_T^2k^2<<\Omega^2$ limit,  to two families of mixed inertial-Tkachenko waves:
\beq
\omega^2&\approx&4\Omega^2(\cos\theta)^2+(1-x_\p)c_T^2k^2(\sin\theta)^4\\
\omega^2&\approx&4\Omega^2(\cos\theta)^2\left(\frac{x_\p}{\varepsilon_\n-x_\p}\right)^2+c_T^2k^2\frac{x_\p^2}{(\varepsilon_\n-x_\p)}(\sin\theta)^4
\eeq
In the limit $x_\p\longrightarrow 1$ and $\varepsilon_\n\longrightarrow 0$ one has again $\varepsilon_\p=\varepsilon_\n=0$ from the relation $\varepsilon_\p=\varepsilon_\n (1-x_\p)/x_\p$, the two fluids decouple and we have the two separate families of modes of equations (\ref{mod1}) and (\ref{mod2}).

\subsection{Mutual friction}

We now consider the dissipative terms due to mutual friction, i.e. to the drag parameter $\mathcal{R}$.  In order to keep the results tractable we take $\varepsilon_\n=\varepsilon_\p=0$.
It is still however impractical to consider the whole solution, so let us first of all consider modes propagating along the $z$ axis ($\theta=0$).
In this case one has two families of inertial modes, one which is undamped with dispersion relation
\be
\omega=\pm 2\Omega
\label{inertia1}
\ee
and one which is affected by mutual friction:
\be
\omega=\pm 2\Omega\tilde{\mathcal{B}}^{'}-i2\Omega\tilde{\mathcal{B}}
\label{inertia2}
\ee
where we remind the reader that $\tilde{\mathcal{B}}^{'}=1-\mathcal{R}^2/[x_\p(1+\mathcal{R}^2)]$ and $\tilde{\mathcal{B}}=\mathcal{R}/[x_\p(1+\mathcal{R}^2)]$. The results in (\ref{inertia1}) and (\ref{inertia2}) agree well with those of \cite{fmode}, in which the authors show that there is one class of inertial modes that corresponds to the fluids co-moving and is undamped (in the absence of chemical coupling) and another class of counter-moving modes that is rapidly damped by mutual friction.

\begin{figure}
\centerline{\includegraphics[height=5.5cm,clip]{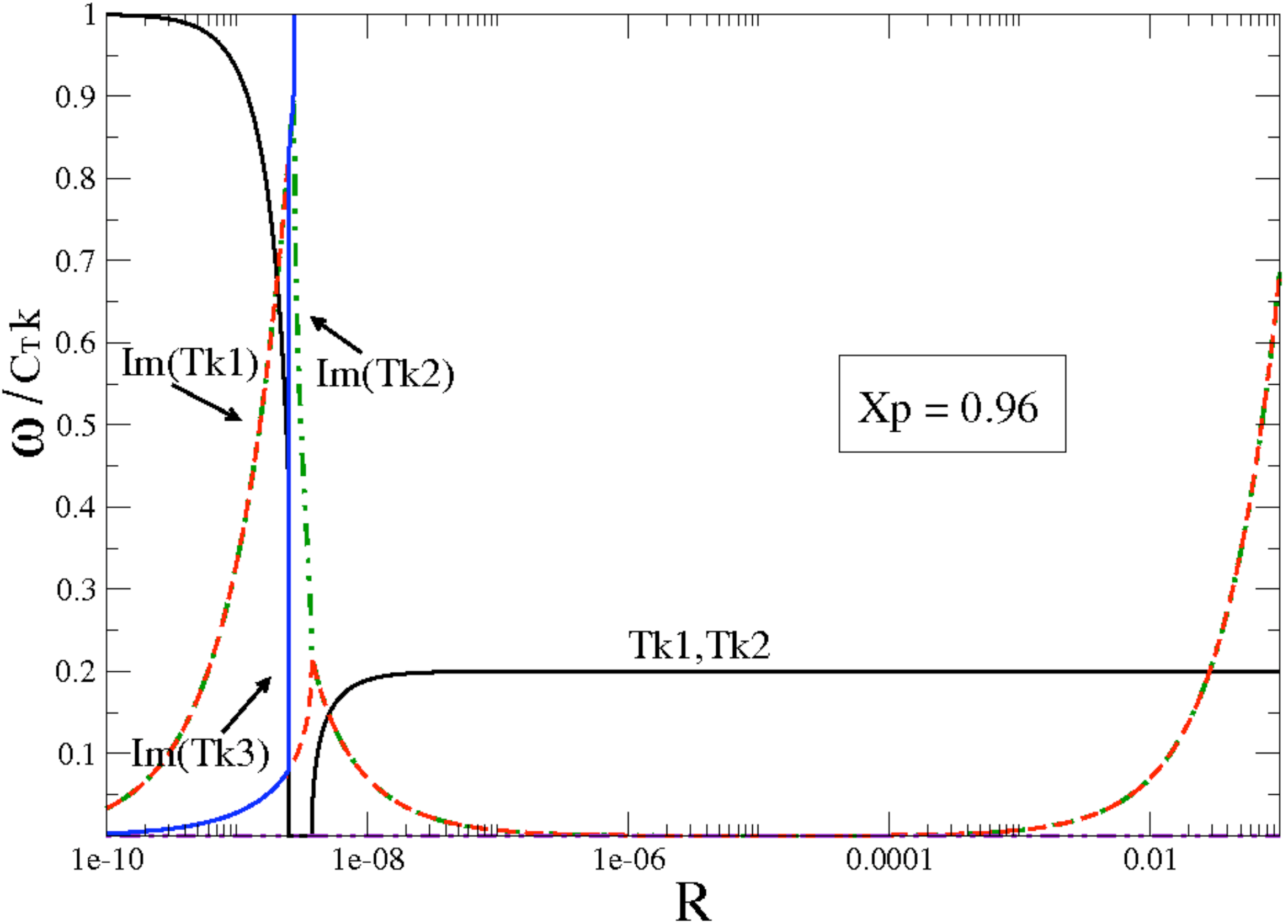}\includegraphics[height=5.5cm,clip]{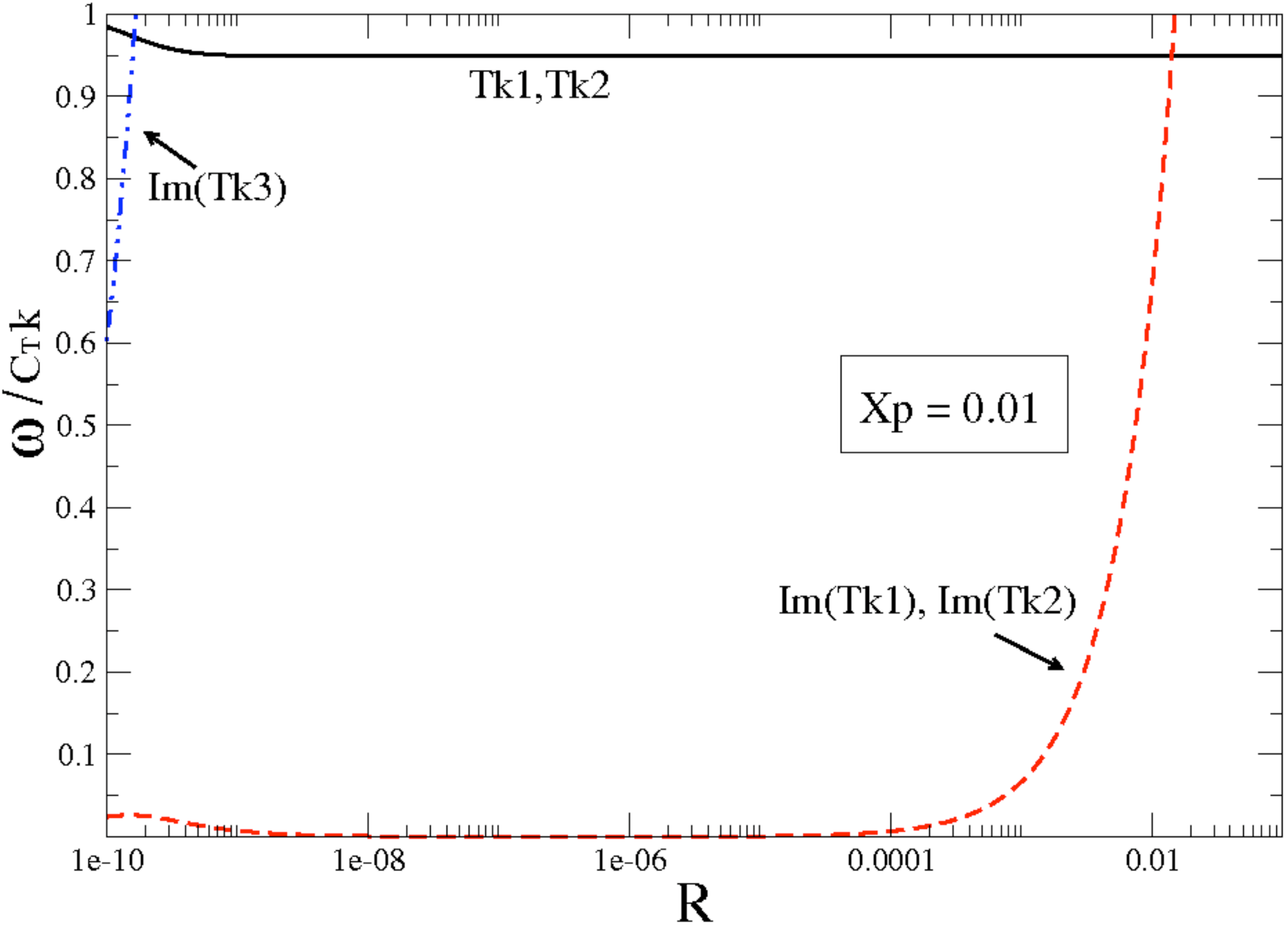}}
\caption{We plot the real part of the modes and the modulus of the imaginary part (dotted lines), for $x_\p$=0.96 and $x_\p$=0.1. We take $k=10^{-6}$. For $x_\p=0.96$ we recover the results of \cite{Sed1}: the real part of the mode vanishes when the mutual friction damping timescale is close to the mode period, and there is an extra purely imaginary root. For the more realistic, but still large, value of $x_\p$=0.1 we see that, on the other hand, the mode is always oscillatory, although the imaginary part is larger when the damping timescale and mode period are similar. For higher values of the drag the frequency of the mode is reduced to $\approx 25\%$ of the original value.}
\label{res1}
\end{figure}

Let us now examine the case of Tkachenko waves propagating perpendicular to the rotation axis ($\cos\theta=0$). In figure \ref{res1} we plot the frequency of the modes for $k=10^{-6}$ and $\nu_{\mbox{star}}=10$ Hz, as a function of $\mathcal{R}$ in the weak drag regime. For large values of the proton fraction $x_\p$ we recover the solution of \cite{Sed1}, in which the real part of the frequency vanishes and the damping becomes large for values of the drag parameter such that the mutual friction damping timescale $\tau_m\approx 1/2\Omega\mathcal{R}$ is approximately equal to the the Tkachenko wave period $P_T=2\pi/\omega_T$ with $\omega_T=k\sqrt{\kappa\Omega/\pi}$. In fact close to this value the mode has an extra purely imaginary root, as found by \cite{Sed1}. For more realistic values of the proton fraction we see, however, that the pathological behaviour disappears and even though the damping is stronger when the mutual friction timescale is close to the period of the modes, the real part does not vanish, the mode is always oscillatory, and there are always only two purely imaginary roots.

\subsection{Perfect pinning}

Up to now we have assumed that the vortex lines are free to move and experience a drag force as they do so. However it is commonly believed that vortex lines can interact strongly with lattice points in the neutron star crust and 'pin' to them, in such a way that they are forced to move with the charged components of the star \citep{pin1,pin2,pin3,pin4,pin5}. The nature of such a pinning force is well beyond the scope of this paper, but to study the propagation of Tkachenko waves in this scenario it is sufficient to consider an unspecified force $f^{\pn}$ acting on the vortices such that they are forced to move with the proton fluid, i.e. such that 
\be
\delta v_\v^i=\delta v_\p^i
\label{velo}
\ee
 In this case the equation of force balance for vortices takes the form:
\be
\rho_\n\kappa n_\v \epsilon_{ijk}\hat{\kappa}^j(\delta v_\p^k-\delta v_\n^k)-\rho_\n{\sigma_i}+f^{\pn}=0\label{pinnedv}
\ee
where there is no drag force acting, as the vortex lines flow with the protons.  If we now consider the neutron and proton fluid there will be a reaction force $-f^{\pn}$ acting on a the protons and the Magnus force acting on the neutrons. Making use of equation (\ref{pinnedv}) we can thus cast the Euler equations in the form:
 \beq
&&\partial_t (\delta v_i^\n + \varepsilon_\n \delta w_i^{\p\n}) + 2\epsilon_{ijk} \Omega^j \delta v_\n^k +\nabla_i \delta \tilde{ \mu}_\n = -2\Omega\epsilon_{ijk} \hat{\kappa}^j(\delta v_\p^k-\delta v_\n^k)\\
&&\partial_t (\delta v_i^\p - \varepsilon_\p \delta w_i^{\p\n}) + 2\epsilon_{ijk} \Omega^j \delta v_\p^k +\nabla_i \delta \tilde{ \mu}_\p = 2\Omega \frac{(1-x_\p)}{x_\p}\epsilon_{ijk}\hat{\kappa}^j(\delta v_\p^k-\delta v_\n^k)-\frac{(1-x_\p)}{x_\p}\sigma_i \label{eulersp}
\eeq
which lead to:
\beq
&&-i\omega \bar{v}_i^\n(1 - \varepsilon_\n) -i\omega\varepsilon_\n bar{v}_i^\p+ 2\epsilon_{ijk} \Omega^j \bar{v}_\n^k +i k_i \bar{ \mu}_\n = -2\Omega\epsilon_{ijk} \hat{\kappa}^j(\bar{v}_\p^k-\bar{v}_\n^k)\\
&&-i\omega\bar{v}_i^\p(1 - \varepsilon_\p)-i\omega\varepsilon_\p \bar{v}_i^\n + 2\epsilon_{ijk} \Omega^j \bar{v}_\p^k +i k_i \bar{ \mu}_\p = 2\Omega \frac{(1-x_\p)}{x_\p}\epsilon_{ijk}\hat{\kappa}^j(\bar{v}_\p^k-\bar{v}_\n^k)-\frac{(1-x_\p)}{x_\p}{\sigma}_i \label{eulersp2}
\eeq
The characteristic equation given by the equations in (\ref{eulersp2}), together with the condition in (\ref{velo}) can be obtained by calculating the determinant of the matrix $K_{ij}$ given in (\ref{ap2}-\ref{ap22}). This leads to two families of modes:
\beq
\omega^2&=&=4\Omega^2(\cos\theta)^2\left(\frac{1-x_\p}{x_\p}\right)^2+c_T^2k^2(\sin\theta)^4\frac{(1-x_\p)^2}{x_\p}\\
\omega^2&=&=4\Omega^2(\cos\theta)^2+c_T^2k^2(\sin\theta)^4(1-x_\p)
\eeq
These are once again mixed inertial-Tkachenko waves, but we see that in the limit $x_\p\longrightarrow 1$ the Tkachenko waves disappear and we are left with only one family of inertial modes. This resembles the situation in superfluid $^4$He, in which even a small amount of pinning swamps the contribution due to lattice elasticity and transforms the Tkachenko waves into inertial waves \cite{Son11}.

\section{ Compressible neutron star matter}

It is well known from the study of superfluid $^4$He that compressibility can have a drastic effect on the mode structure \cite{Reatto}. Including compressibility in the equations of motion for the superfluid leads to a dispersion relation of the form \cite{Son11}:
\be
\omega^2=\pm\frac{c^2_Tc^2_sk^4}{4\Omega^2+c^2_sk^2}
\ee
where $c_s$ is the sound speed. In the long wavelength limit ($k<<\Omega/c_s$) the nature of the mode is thus profoundly altered and the dispersion relation is no longer linear in $k$, but rather parabolic, leading to the so-called "soft" Tkachenko wave frequency:
\be
\omega\approx \pm \frac{c_Tc_s}{2\Omega}k^2
\ee
In the study of $^4$He the long wavelength limit is, however, mainly of theoretical interest, as one would need containers of several hundreds of meters in diameter to explore it experimentally. The situation is very different for BECs as, in contrast with a strongly interacting Bose liquid such as $^4$He, they are weakly interacting Bose gases with low sound speeds for which the effect of compressibility is important at high rotation rates. For BECs the compressible Tkachenko wave spectrum has thus been studied both theoretically \cite{Baym1, bectheor, Baym2} and experimentally \cite{becex}. 

Let us now consider a realistic neutron star . The situation is clearly quite complex as not only can we be in the long wavelength limit ($\Omega\approx c_\s k$) for the more rapidly rotating pulsars, but one also has to account for multi-fluid effects and chemical coupling between the different constituents via the equation of state.
One cannot, in general, assume incompressibility for the proton and neutron fluids and it is clearly of great interest to adapt our formalism to include the effects of compressibility and chemical coupling.
To study this problem it is now advantageous to write the perturbation equations in the form of equations (\ref{tutto1})-(\ref{tutto3}). In the plane wave approximation the Euler equations take the form (in the Cowling approximation):
\beq
&&-i\omega\bar{v}_i +i \frac{k_i}{\rho}\bar{p}-\frac{\bar{\rho}}{\rho}\nabla_i p +2\epsilon_{ijk}\Omega^j\bar{v}^k=-(1-x_\p)\sigma_i\\
&&-i\omega(1-\bar{\varepsilon})\bar{w}_i+i k_i\bar{\beta}+2\epsilon_{ijk}\Omega^j\bar{w}^k=-2\Omega\frac{\mathcal{R}}{x_\p} (i\omega\bar{\epsilon}_i+\bar{v}_i+(1-x_\p)\bar{w}_i)+\sigma_i
\label{Eulkt1}
\eeq
and the continuity equations can be written as:
\beq
&&-i\omega\bar{\rho}+i\rho k_j\bar{v}^j+\bar{v}^j\nabla_j\rho=0\\
&&-i\omega\bar{x}_\p+i x_\p(1-x_\p) k_j\bar{w}^j+\bar{w}^j\nabla_j[\rho x_\p(1-x_\p)]=0,\label{Contkt1}
\eeq
while the equation of force balance still takes the form in (\ref{forceb}). As we are now considering compressible matter, we will also need an equation of state for the perturbations.
Choosing to work with the density ($\bar{\rho}$) and proton fraction ($\bar{x}_\p$) perturbations, one can write:
\beq
\bar{p}&=&\left(\frac{\partial p}{\partial \rho}\right)\bar{\rho}+\left(\frac{\partial\rho}{\partial x_\p}\right)\bar{x}_\p\\
\bar{\beta}&=&\left(\frac{\partial \beta}{\partial \rho}\right)\bar{\rho}+\left(\frac{\partial\beta}{\partial x_\p}\right)\bar{x}_\p
\label{eos}
\eeq
Ideally the partial derivatives of the thermodynamical variables in equation (\ref{eos}) should be derived from a fully consistent multi-parameter equation of state, which should also allow us to calculate the entrainment parameters and the superfluid gaps for neutrons and protons. However, not only is such an equation of state not currently available, but its use would also be beyond the scope of our simplified plane wave analysis.
In order to keep the problem tractable we shall assume that our background model is described by an $n=1$ polytrope and use for the perturbations two simplified analytic equations of state, that are essentially extensions of a single-fluid $n=1$ polytrope. 

First of all we shall consider the equation of state of \cite{rmode}, which we refer to as model A. In this case we have:
 \beq
 &&\left(\frac{\partial p}{\partial \rho}\right)=c_s^2,\;\;\;\;\;\;\left(\frac{\partial p}{\partial x_\p}\right)=\frac{\rho c_s^2}{x_\p}\\
 &&\left(\frac{\partial \beta}{\partial \rho}\right)=\frac{c_s^2}{\rho x_\p},\;\;\;\;\;\;\left(\frac{\partial \beta}{\partial x_\p}\right)=\frac{c_s^2}{x_\p^2}
  \eeq
  where $c_s$ is the sound speed of the background 
We shall then consider a second model, in order to understand the importance of the chemical coupling on the mode structure. This model, which we refer to as model B, is essentially a re-parametrisation of the model II equation of state in \cite{PR} (also used in \cite{Andrea} where it is denoted as model B0) and takes the form:

\beq
 &&\left(\frac{\partial p}{\partial \rho}\right)=c_s^2,\;\;\;\;\;\;\left(\frac{\partial p}{\partial x_\p}\right)=\alpha \frac{\rho c_s^2}{x_\p}\\
 &&\left(\frac{\partial \beta}{\partial \rho}\right)=\alpha \frac{c_s^2}{\rho x_\p},\;\;\;\;\;\;\left(\frac{\partial \beta}{\partial x_\p}\right)=\gamma \frac{c_s^2}{x_\p^2}
\eeq 
 We can thus study the behaviour of the solutions to are problem as we vary the parameters $\alpha$ and $\gamma$. Clearly model A corresponds to the case $\alpha=1$, $\gamma=1$.
 
 The sound speed in the background, for an $n=1$ polytrope, takes the form
  \be
  c_s=2K\rho
  \ee
  where $K=2GR^2/\pi$ depends only on the stellar radius.
 However in our plane wave approximation we shall assume that the background quantities vary over a length-scale greater than that of the oscillations, and thus take them to be constant and neglect their gradients. This approximation is not necessarily justified, as in the crust the density and pressure vary by several orders of magnitude over a length-scale of approximately 1 km, which is comparable with the longest wavelengths we consider for our Tkachenko waves. It is, however, a reasonable approximation for shorter wavelengths and in the neutron star core. 
The sound speed will thus be a constant in our formulation and specifically we take $c_s=10^9$ cm s$^{-1}$. We also take the proton fraction, which in a rigorous description should also be derived from the equation of state, as a constant and will study the effect that varying it can have on the modes.
Needless to say, future work should aim to relax this approximation and consider a fully stratified neutron star. 

Finally let us remark that for simplicity we take $\bar{\varepsilon}=0$ in the following discussion. We have experimented with varying the parameter $\bar{\varepsilon}$ between -0.8 and 0.8, but it is found to have very little effect on the dispersion relation.

\subsection{Undamped propagation}

\subsubsection{Model A}

Let us consider, first of all, the undamped propagation of waves in a neutron star. We thus take $\mathcal{R}=0$ and, to keep the problem tractable, $\bar{\varepsilon}=0$, and solve the characteristic equation obtained form the determinant of the matrix $K_{ij}$ given in (\ref{ap3}). 
As a first step we focus on model A. The results are two families of modes, such that:
\beq
\omega^2&=&4\Omega^2- \frac{c_T^4k^4(\sin\theta)^4}{4\Omega^2}\label{cacio1}\\
\omega^2&=&\pm\frac{1}{2}\left(4\Omega^2 + {k^2 c_s^2}\right)+\frac{1}{2x_\p}\sqrt{(k^2c_s^2+4\Omega^2x_\p)^2-x_\p(4\Omega k c_s\cos\theta)^2}
\eeq
As we can see we obtain sound waves and rotationally corrected Tkachenko waves, for which the main contribution to the frequency is now given by the stellar rotation frequency. This result is somewhat surprising, as in this limit the classical Tkachenko waves no longer exist and the vortex elasticity simply provides a small correction to what is, in essence, the frequency of an inertial wave. In this case, for a pulsar rotating at $\approx 1$Hz, the frequency of the Tkachenko waves would be much too high to explain the observed periodicities of $100\sim 1000$ days observed in the timing residuals.
Such a drastic modification in the dispersion relation clearly needs to be investigated in more detail. We have, after all, used a simplified version of the equation of state, so let us turn our attention to model B in order to understand how varying the parameters (and thus the coupling between the components) can affect the mode structure and whether there is a reasonable set of parameters for which one can still obtain the usual Tkachenko waves.

\subsubsection{Model B}

\begin{figure}
\centerline{\includegraphics[height=5.5cm,clip]{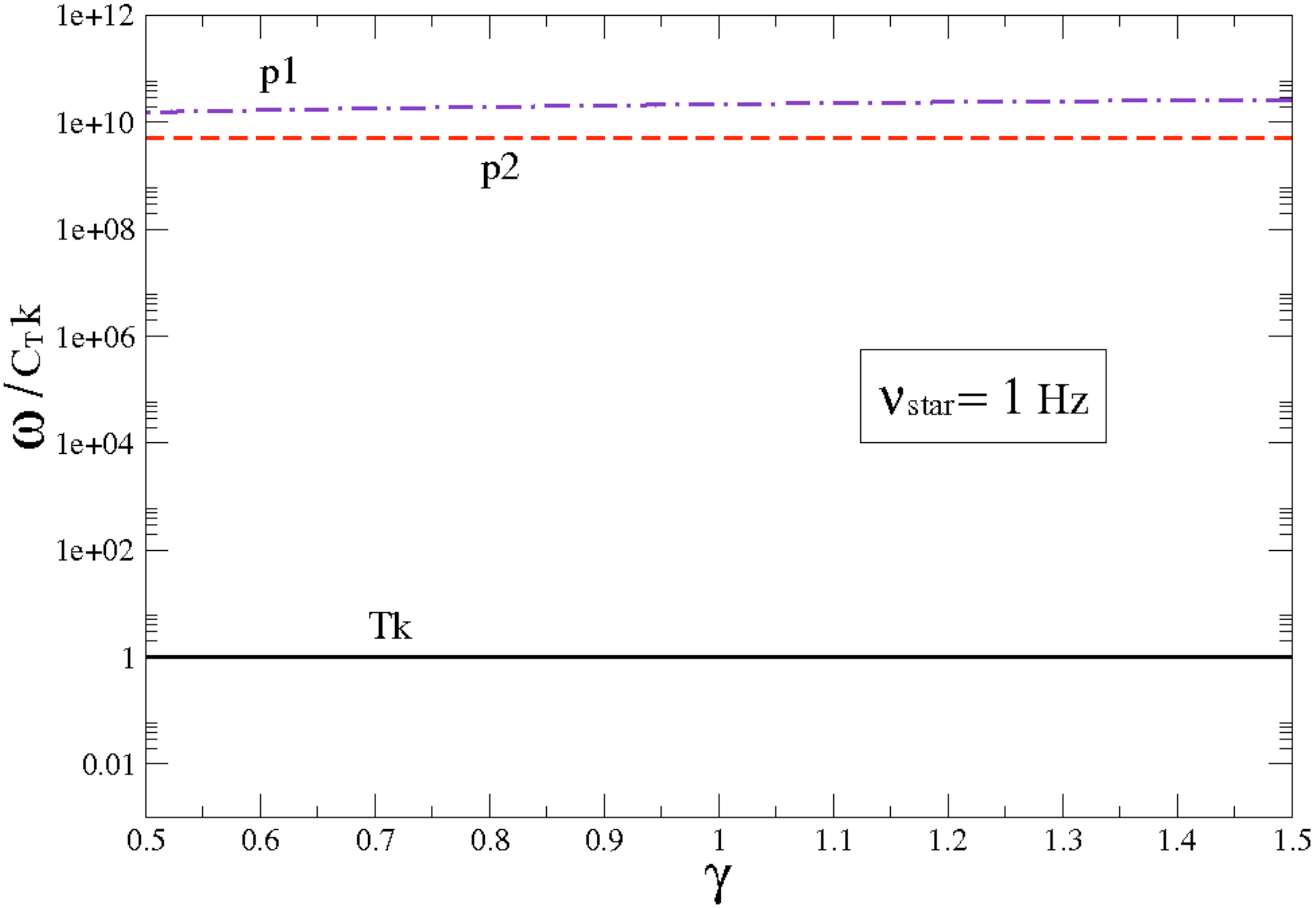}\includegraphics[height=5.5cm,clip]{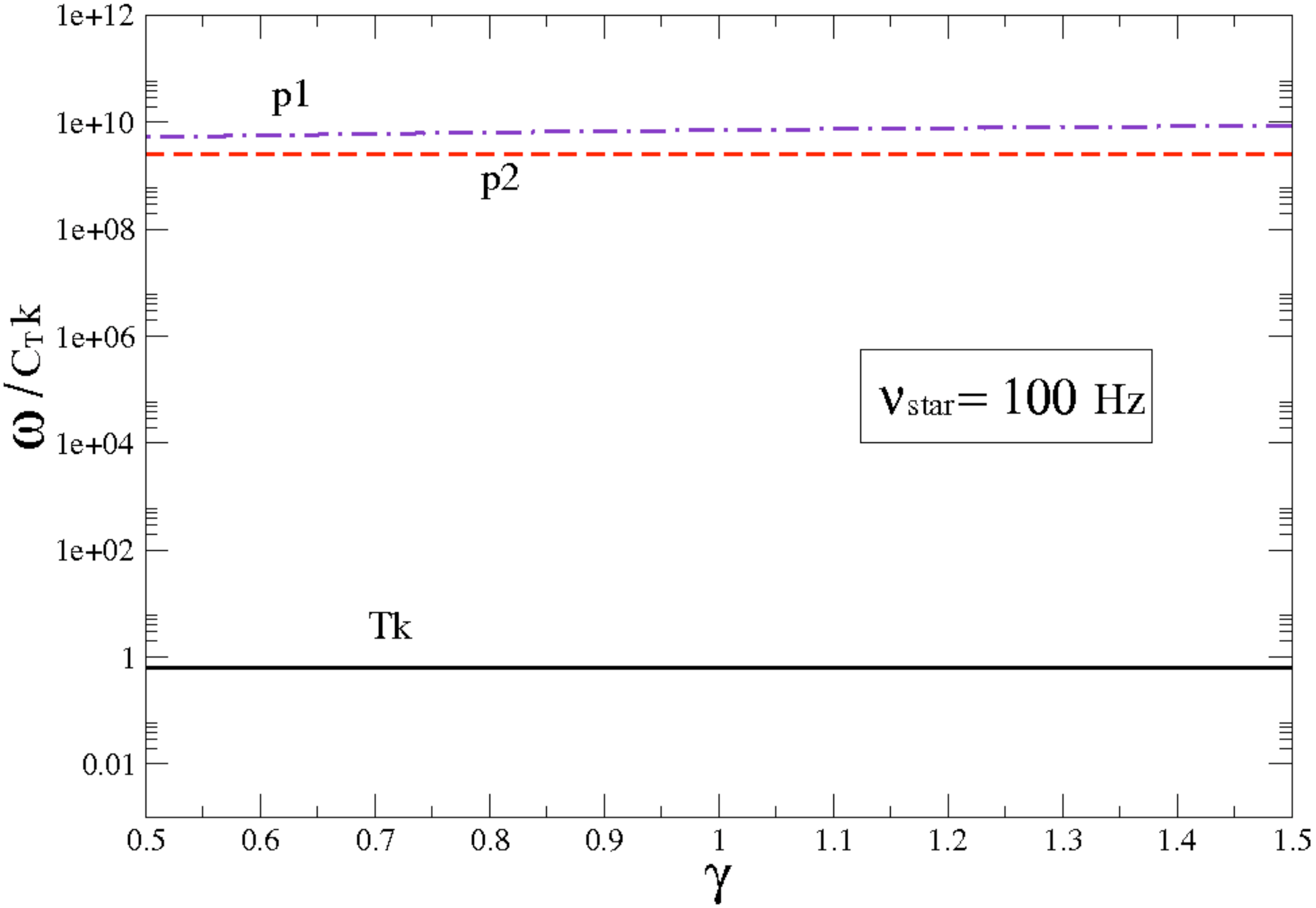}}
\caption{We plot the frequencies of the modes we obtain, normalised to the classical Tkachenko wave frequency, for two different rotation rates of the star and for a varying parameter $\gamma$, whilst keeping $\alpha=0$. We take $x_\p=0.05$. We can see that we have two families of high frequency sound waves and then the Tkachenko waves, the frequency of which is shifted from the classical value at higher rotation rates. This is expected as for higher rotation rates the effects of compressibility become more important. Furthermore it is clear from the graph that, although varying $\gamma$ has a small effect on the frequency of the sound waves, it has no effect on the frequency of the Tkachenko waves. In these plots we have set $\bar{\varepsilon}=0$, $k=10^{-6}$ and taken $\gamma=1.5$.}
\label{variob}
\end{figure}

\begin{figure}
\centerline{\includegraphics[height=5.5cm,clip]{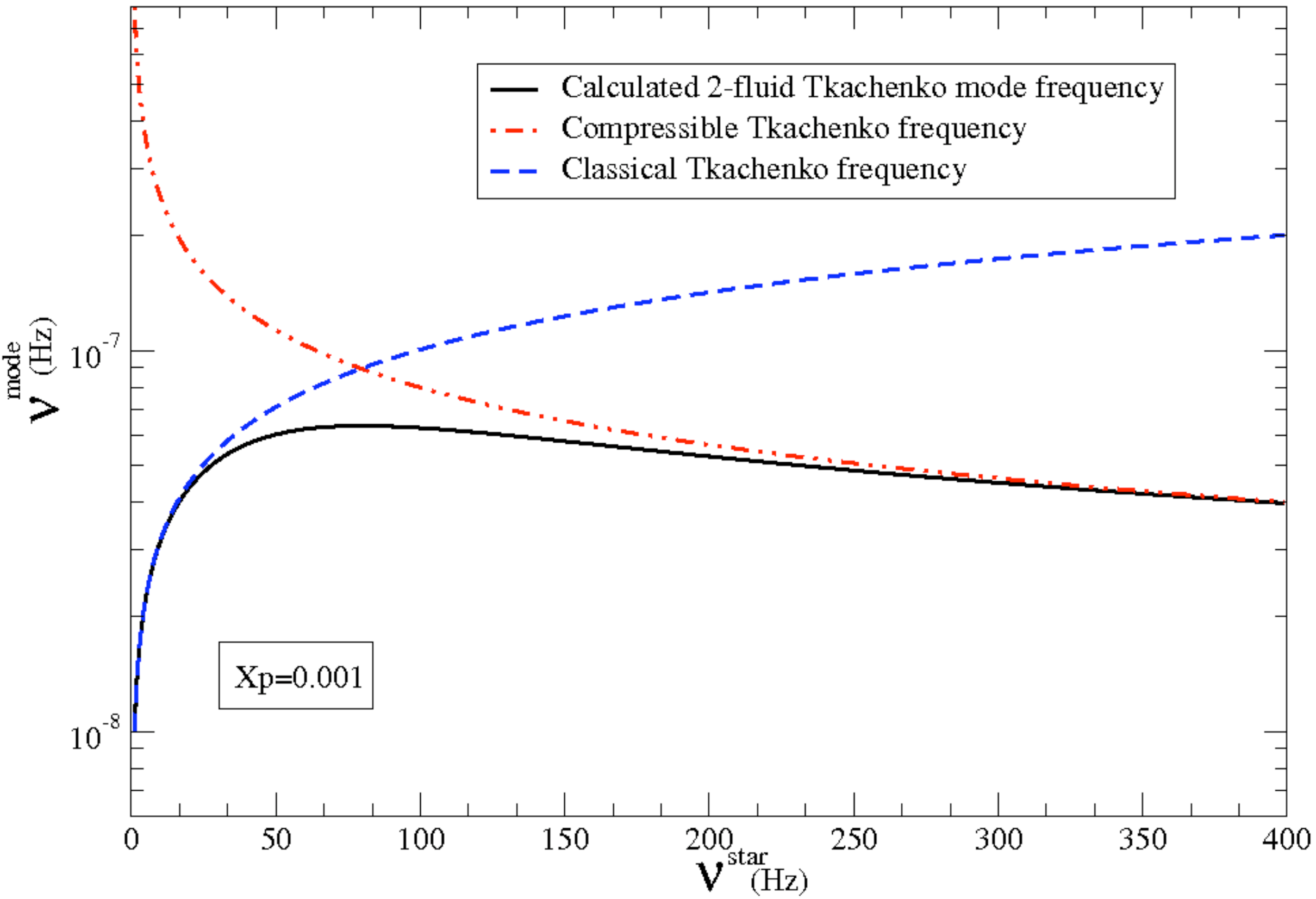}\includegraphics[height=5.5cm,clip]{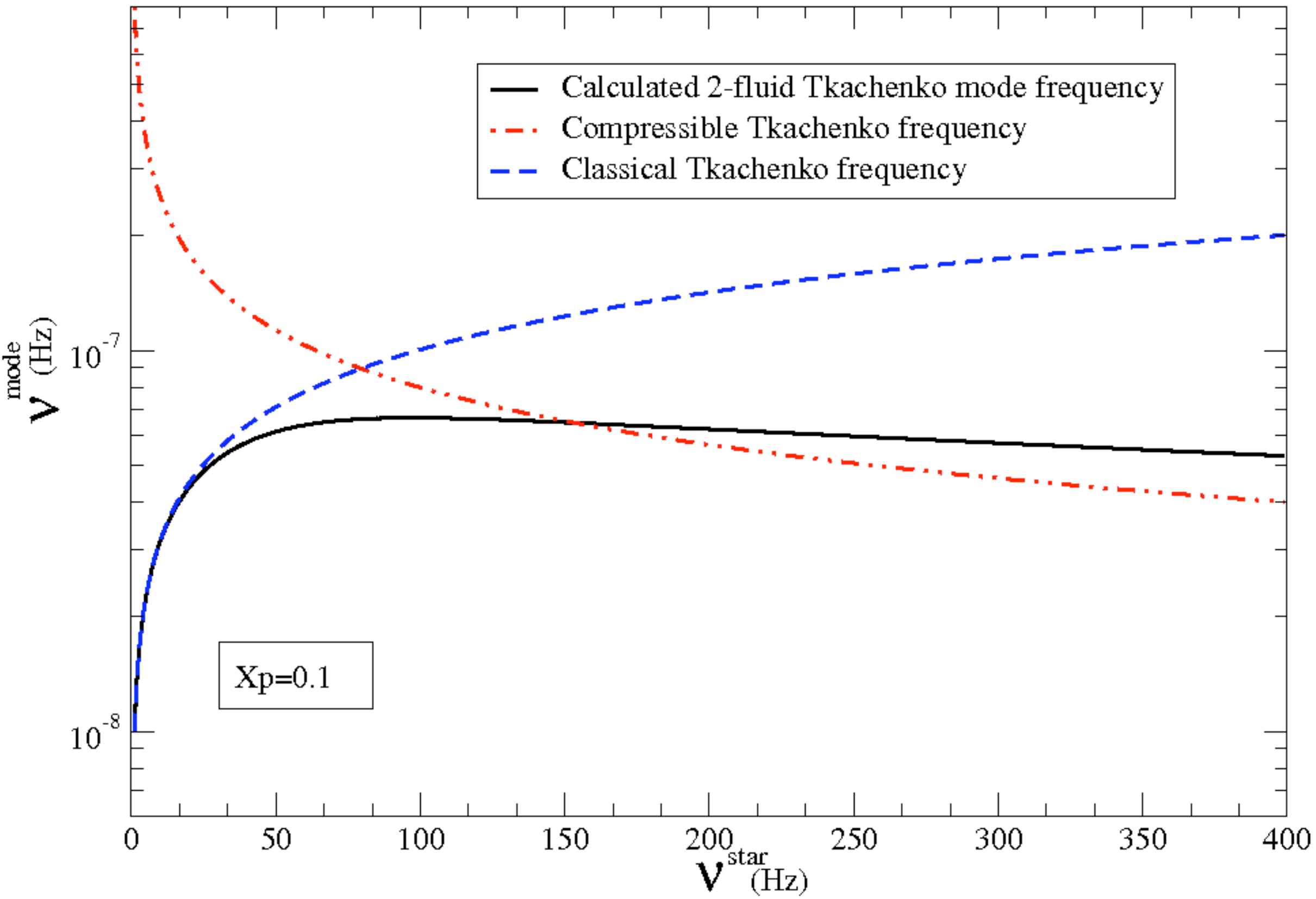}}
\caption{We plot the frequencies of the Tkachenko waves we obtain numerically and compare them to the classical Tkachenko wave frequency $c_T k$ and to the "soft" Tkachenko wave frequency $c_Tc_sk^2/2\Omega$. As we can see the frequency tends to that of the "soft" mode for higher rotation frequencies (in the millisecond range, which is that of the fastest known pulsars) and is slightly modified by multi-fluid effects for high values of $x_\p$. In these plots we have set $\bar{\varepsilon}=0$ and taken $k=10^{-6}$.}
\label{varioom}
\end{figure}

In the case of model B the extra parameters make it necessary to solve the characteristic equation numerically, but they also allow us extra freedom to explore different regimes for the equation of state. First of all let us eliminate all chemical coupling by setting $\alpha=0$. In figure \ref{variob} we plot the mode frequency for varying $\gamma$ at different rotation rates. We see that in this case there is a mode with the frequency of the classical Tkachenko mode and furthermore varying $\gamma$ has very little effect on its frequency. If we now keep $\gamma$ fixed and vary the rotation rate of the star we see in figure \ref{varioom} that for low rotation rates the frequency of the Tkachenko mode is the classical one. For higher rotation rates the classical frequency increases but the frequency of the Tkachenko mode tends to that of the so called "soft" Tkachenko mode $c_Tc_sk^2/2\Omega$. The result is thus that, for longer wave-length of order the stellar radius, there is always a Tkachenko mode with periods consistent with the $100-1000$ day variability typical of pulsar timing noise. 
This picture is clearly very different from that of the previous section, in which the Tkachenko waves had essentially disappeared, so let us investigate how re-instating the chemical coupling and varying the parameter $\alpha$ can affect the mode structure.


\begin{figure}
\centerline{\includegraphics[height=5.5cm,clip]{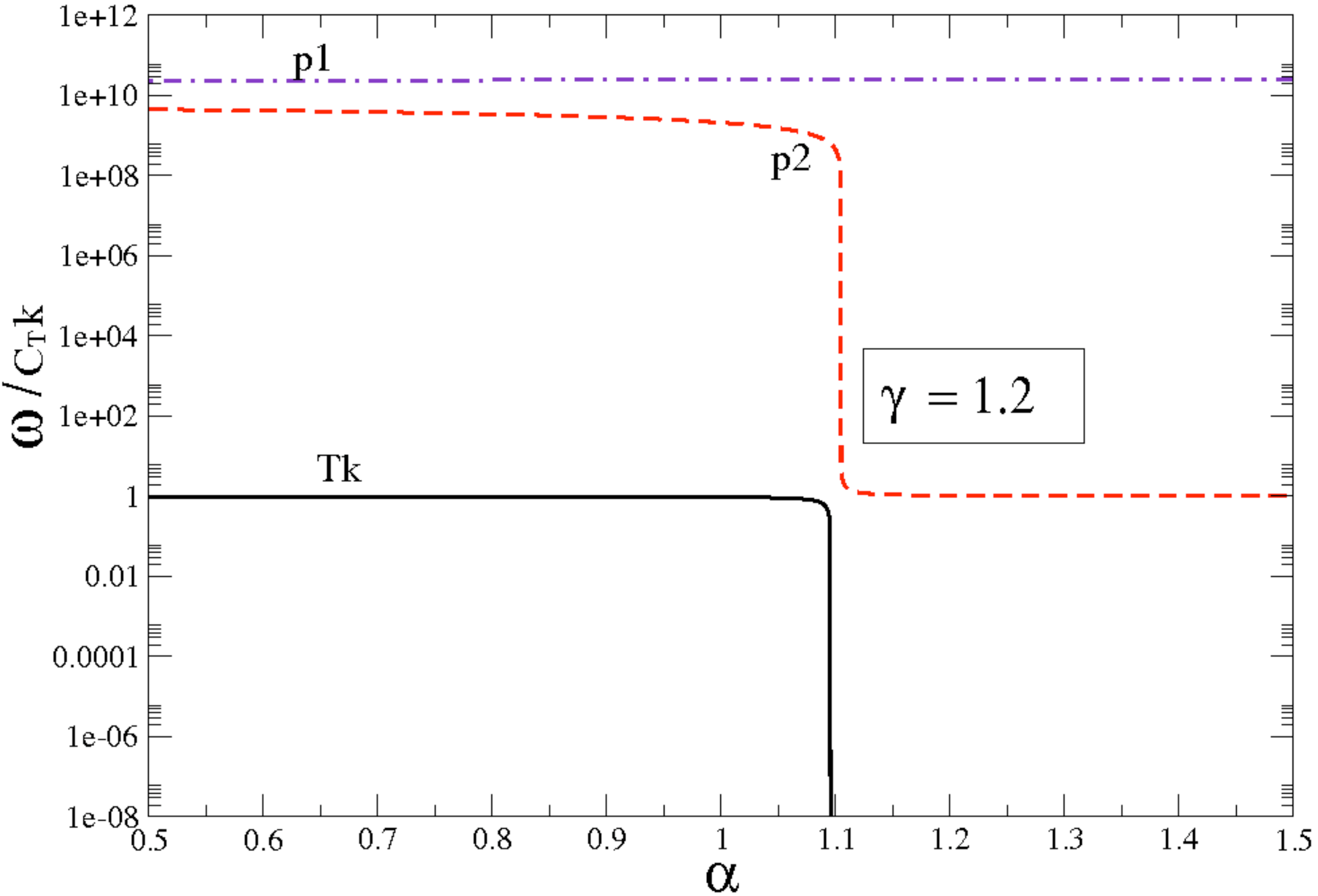}\includegraphics[height=5.5cm,clip]{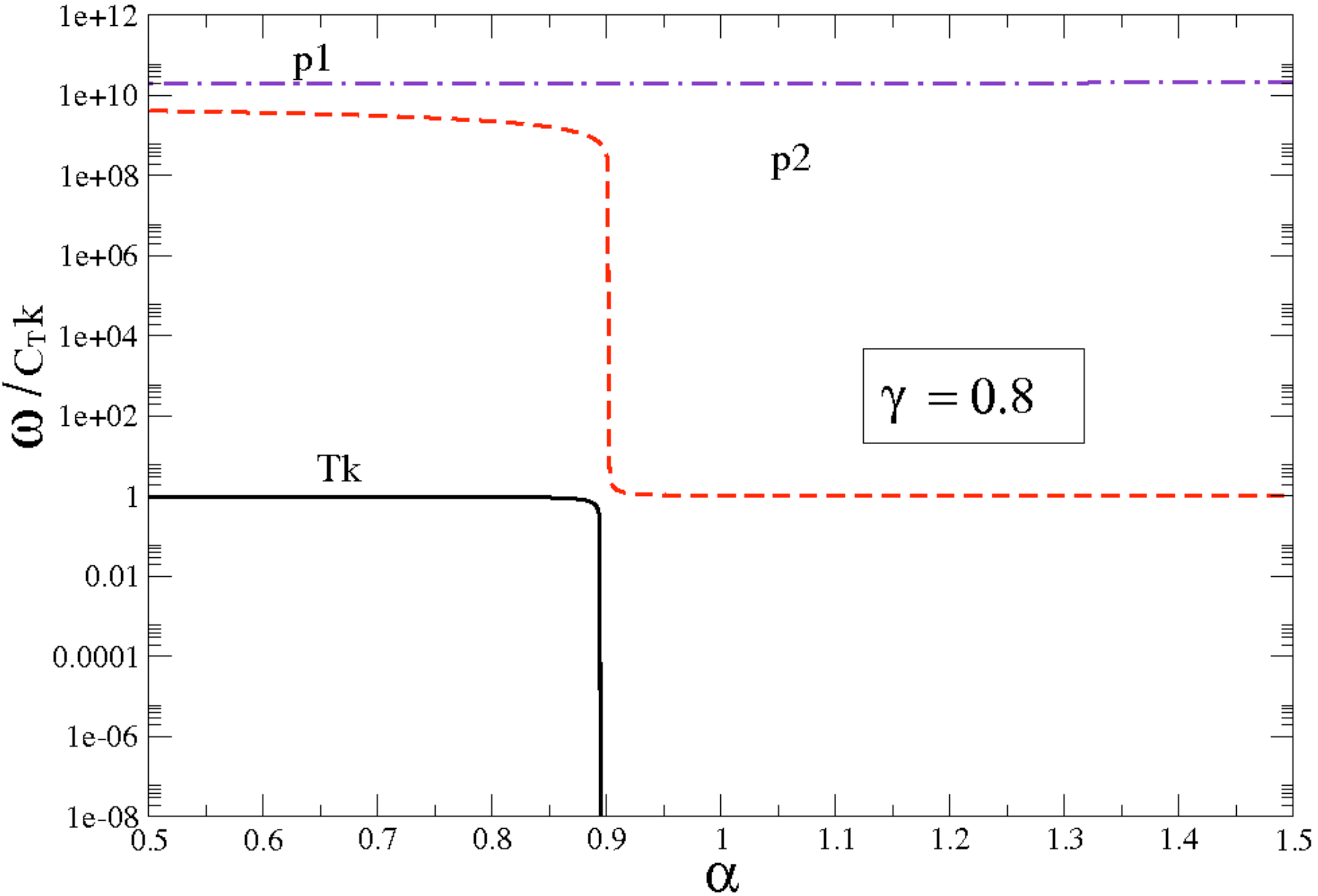}}
\caption{For a stellar rotation rate of 10 Hz, we plot the frequency of the modes, normalised to the classical Tkachenko mode frequency, for varying values of the parameter $\alpha$. We see that there is still a family of sound waves (indicated as $p1$), but there is now an avoided crossing between the second family of sound waves ($p2$) and the Tkachenko waves ($Tk$), with the frequency of the sound wave becoming that of a classical Tkachenko wave. There is thus a vast portion of parameter space in which one has a mode close to the frequency of a classical Tkachenko wave, but there exists a small region where the mode crossing occurs, in which the Tkachenko mode is not oscillatory in nature and the frequency of the sound waves is too high to account for the slow variability of pulsar timing residuals. Once again we have set $\bar{\varepsilon}=0$ and taken $k=10^{-6}$.}
\label{variotutto}
\end{figure}

In figure \ref{variotutto} we plot the mode frequency for a stellar rotation frequency of 10 Hz varying $\alpha$ whilst keeping $\gamma$ fixed. The result is now much more intriguing as one still has one family of sound waves, but there is then an avoided crossing between the second family of sound waves and the Tkachenko waves, with the frequency of the sound wave becoming that of a classical Tkachenko wave as we vary $\alpha$. The real part of frequency of the Tkachenko wave, on the other hand, vanishes after the avoided crossing and one obtains two purely imaginary roots.

\begin{figure}
\centerline{\includegraphics[height=5.5cm,clip]{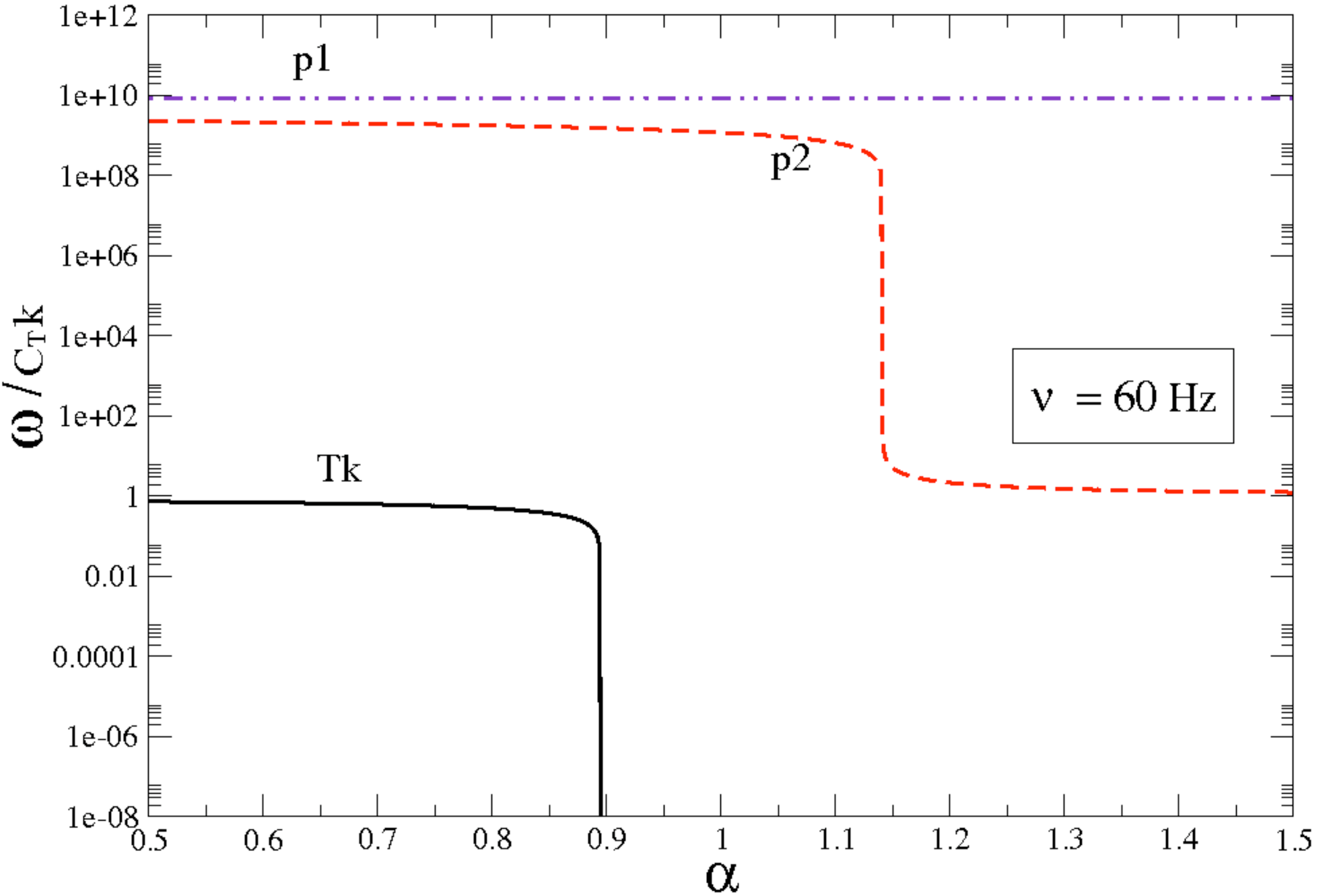}\includegraphics[height=5.5cm,clip]{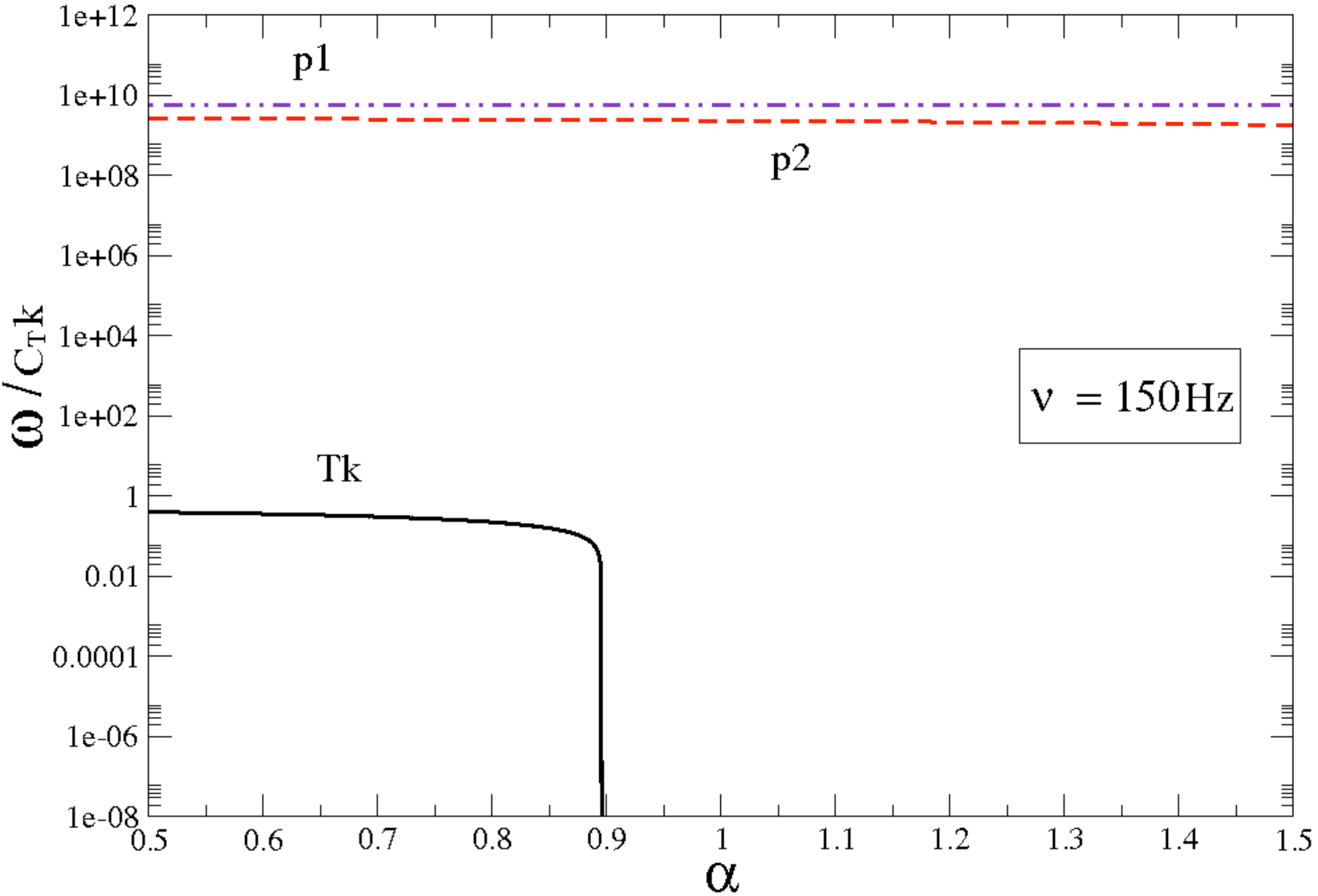}}
\caption{We plot the frequencies of the modes we obtain (sound waves $p1$ and $p2$ and Tkachenko waves $Tk$), normalised to the classical Tkachenko wave frequency, for two different rotation rates of the star and for a varying parameter $\alpha$, whilst keeping $\gamma=0.8$. We take $x_\p=0.05$. We can see that as the rotation rate increases not only does the Tkachenko wave frequency decrease as expected, but there is also a vast region of parameter space in which the Tkachenko mode dissapears. In these plots again we have set $\bar{\varepsilon}=0$ and taken $k=10^{-6}$.}
\label{varioOMT}
\end{figure}

Summarising there is a vast portion of parameter space in which one has a mode close to the frequency of a classical Tkachenko wave but there exists a small region (which thus includes model A for which $\alpha=\gamma=1$) where the avoided crossing occurs, in which the Tkachenko mode is not oscillatory in nature and the frequency of the sound waves is too high to account for the slow variability of pulsar timing residuals.
If we now increase the stellar rotation rate we can see from figure (\ref{varioOMT}) that the region in which the Tkachenko mode is not oscillatory becomes even larger, allowing for vast portions of parameter space in which there is no mode close to the Tkachenko frequency.
However we have not yet considered the impact of mutual friction damping, which could potentially limit even more the range of parameters for which one has long lived Tkachenko oscillations. Let us thus move on to consider the full problem.


\subsection{Mutual  Friction}

\begin{figure}
\centerline{\includegraphics[height=5.5cm,clip]{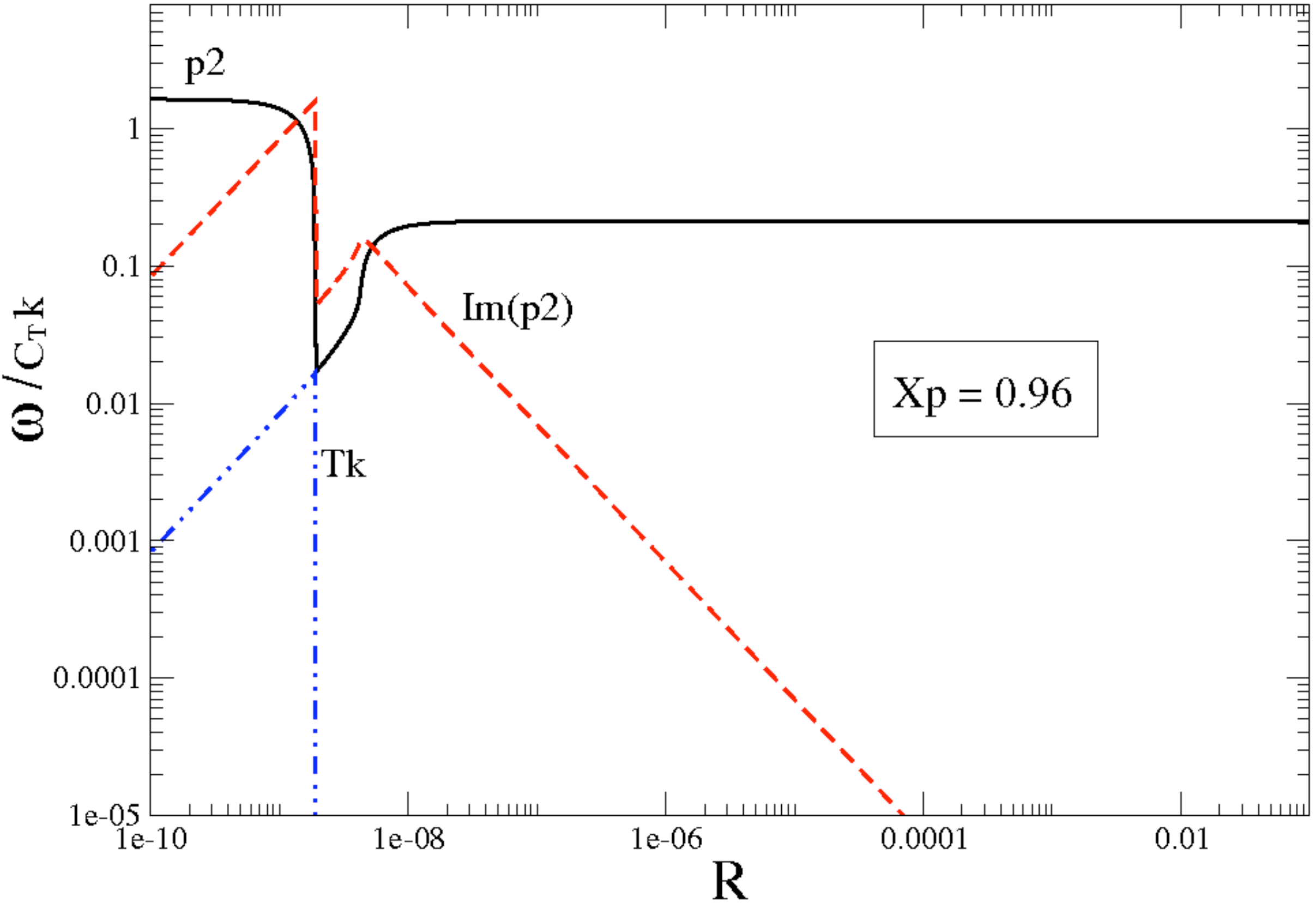}\includegraphics[height=5.5cm,clip]{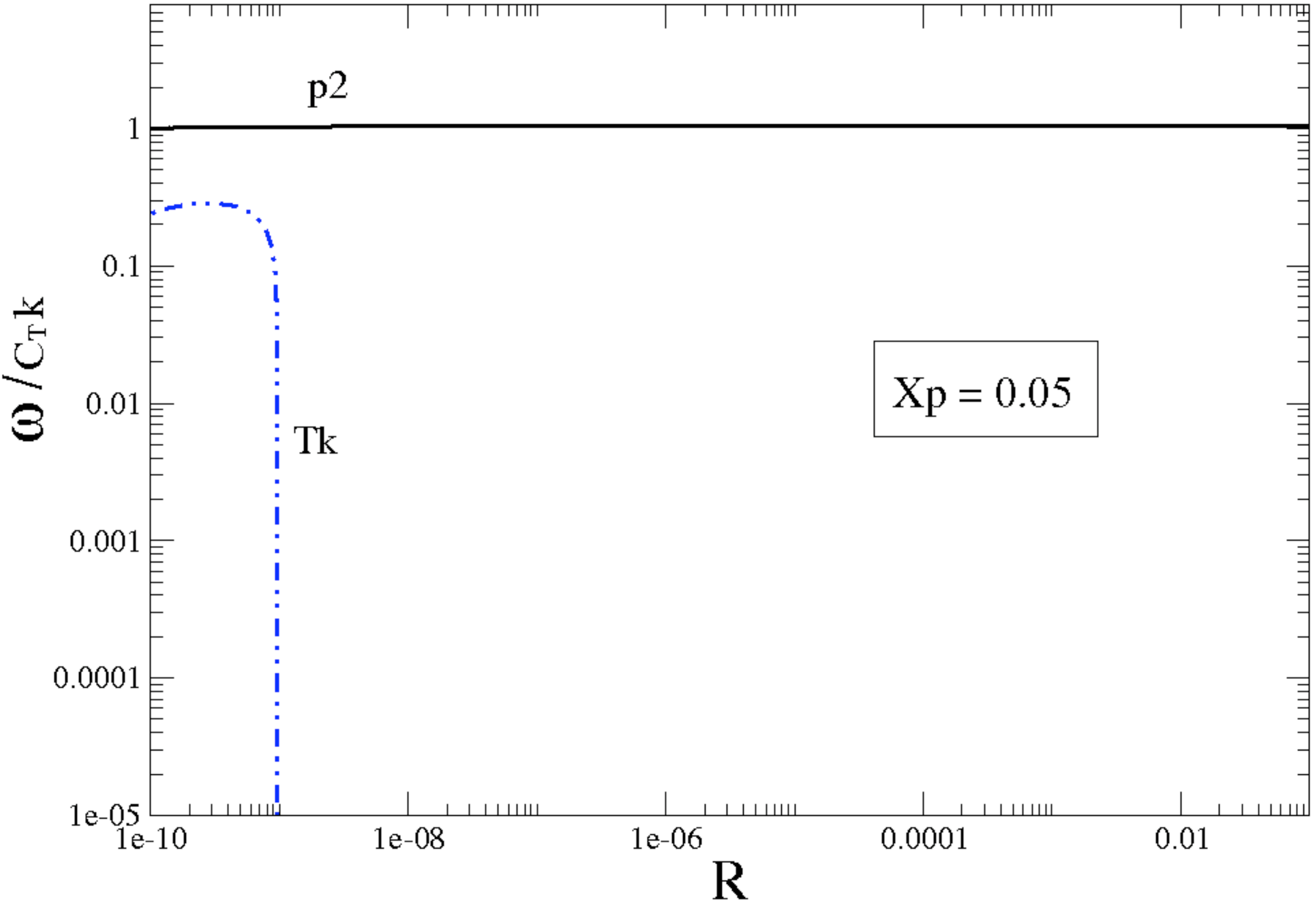}}
\centerline{\includegraphics[height=5.5cm,clip]{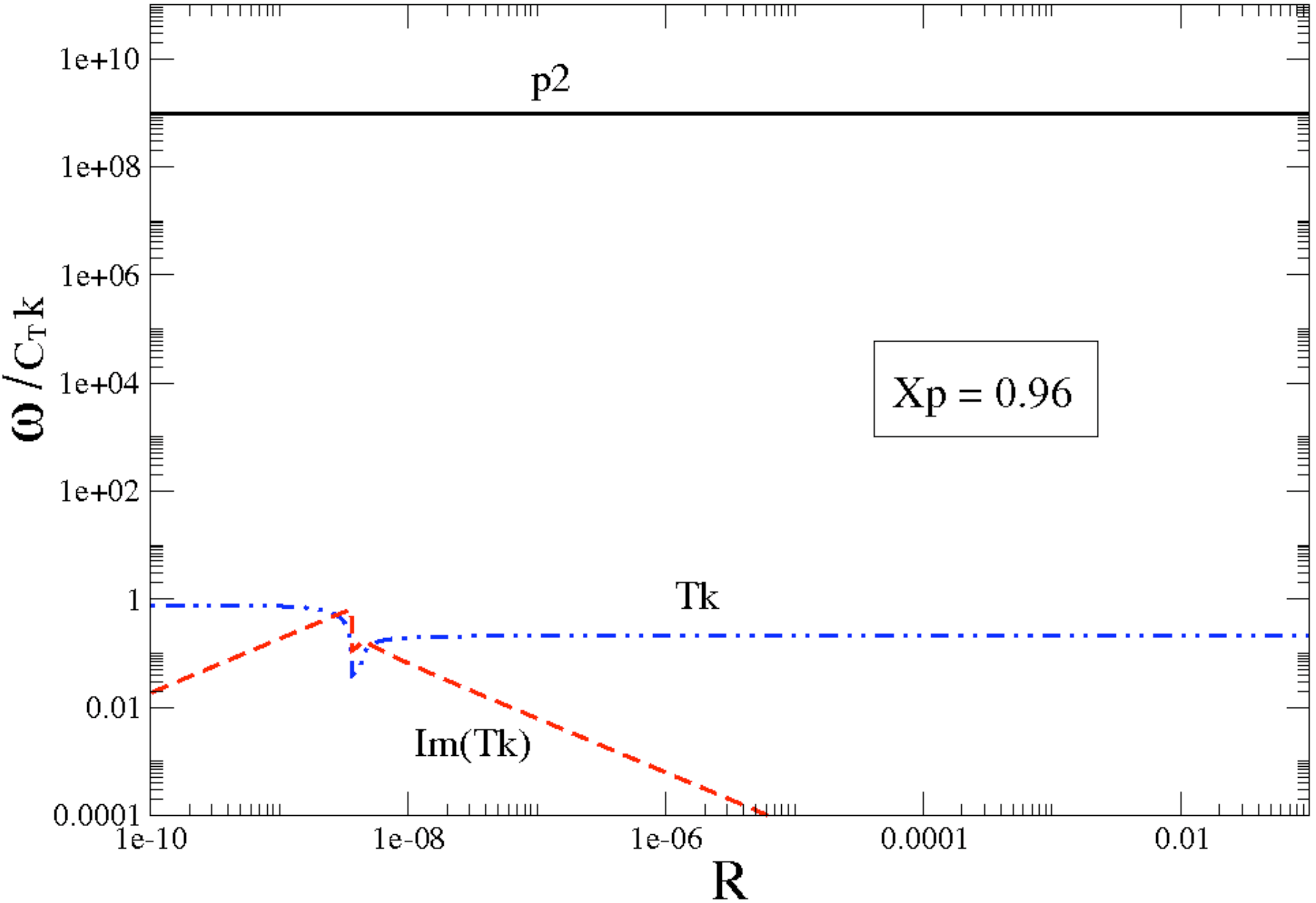}\includegraphics[height=5.5cm,clip]{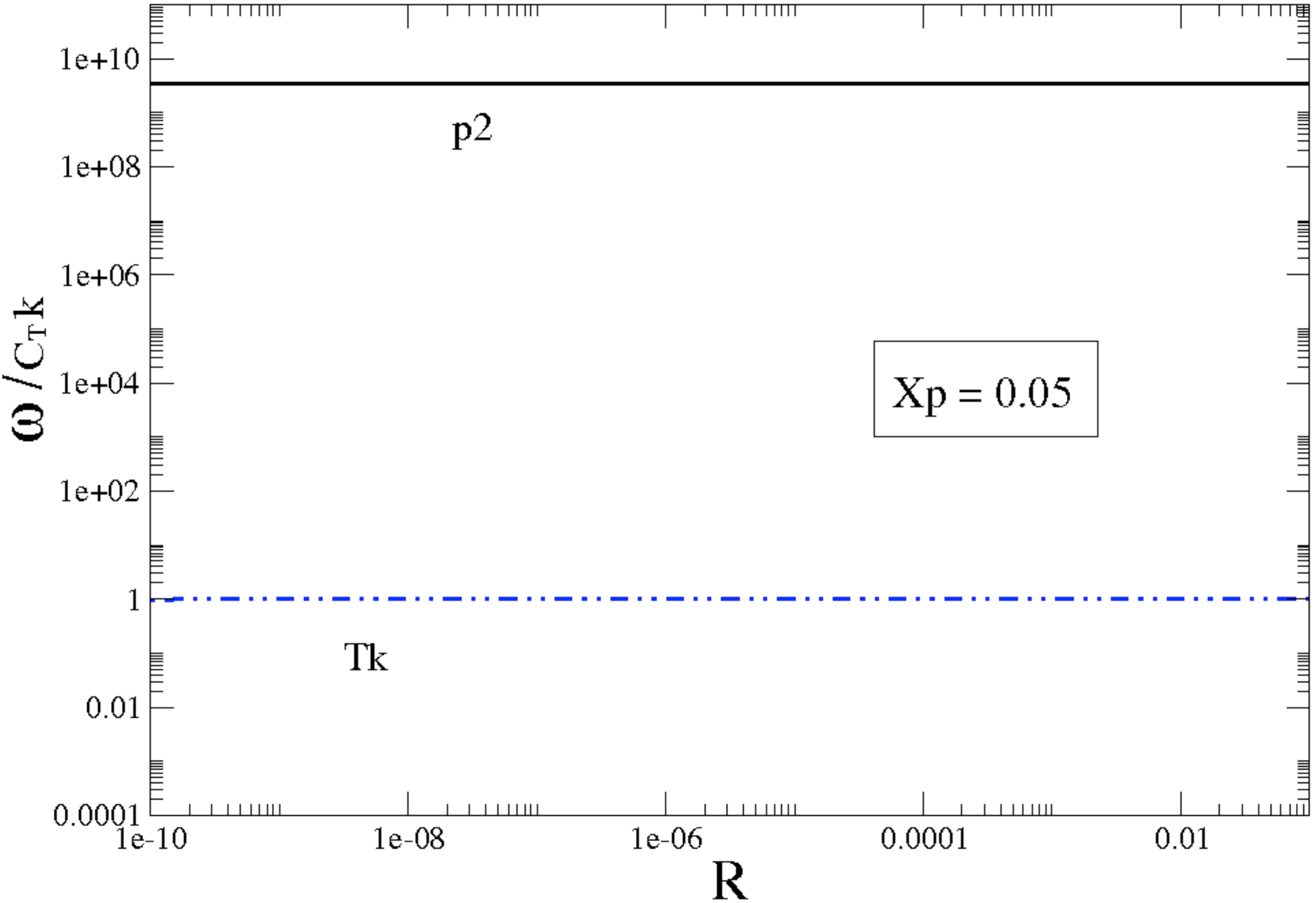}}
\caption{We plot the Tkachenko and second sound waves for a rotation rate of $10$ Hz, for $k=10^{-6}$, $\bar{\varepsilon}=0$ and $\cos\theta=0$. In the top panel we take $\alpha=0.8$ and $\beta=1.2$, in the bottom panel we take $\alpha=0.8$ and $\beta=1.2$.For $\beta>\alpha$ and slow rotation of the star one has Tkachenko waves close to the classical Tkachenko frequency, while for $\beta<\alpha$ one has an avoided crossing between the second family of sound waves and the Tkachenko waves. In both cases the waves oscillating close to Tkachenko frequency are strongly damped by mutual friction in a narrow range of the parameter $\mathcal{R}$ and for large values of $x_\p$. As expected the mutual friction damping is weak in both cases for lower, more realistic values of $x_\p$. }
\label{comparo}
\end{figure}

The inclusion of mutual friction makes the problem considerably more complicated, and once again the characteristic equation must be solved numerically. The picture that emerges is however not very different from that of the previous section. One still finds a high frequency family of sound waves, largely unaffected by mutual friction, then a second family of sound waves at lower frequency and a family of Tkachenko waves. In addition one has two purely imaginary roots to the characteristic equation. 
Let us focus on the Tkachenko waves and on the lower frequency sound waves. As we can see in figure (\ref{comparo}) the behaviour of the modes depends strongly on the chemical coupling. For $\beta>\alpha$ and slow rotation of the star one has Tkachenko waves close to the classical frequency, while for $\beta<\alpha$ one has an avoided crossing between the second family of sound waves and the Tkachenko waves. In both cases the waves oscillating close to Tkachenko frequency are strongly damped by mutual friction in a narrow range of the parameter $\mathcal{R}$ for large values of $x_\p$, exactly as in the incompressible case. For more realistic values of $x_\p$ we find that, as expected, the damping is negligible. 

\begin{figure}
\centerline{\includegraphics[height=5.5cm,clip]{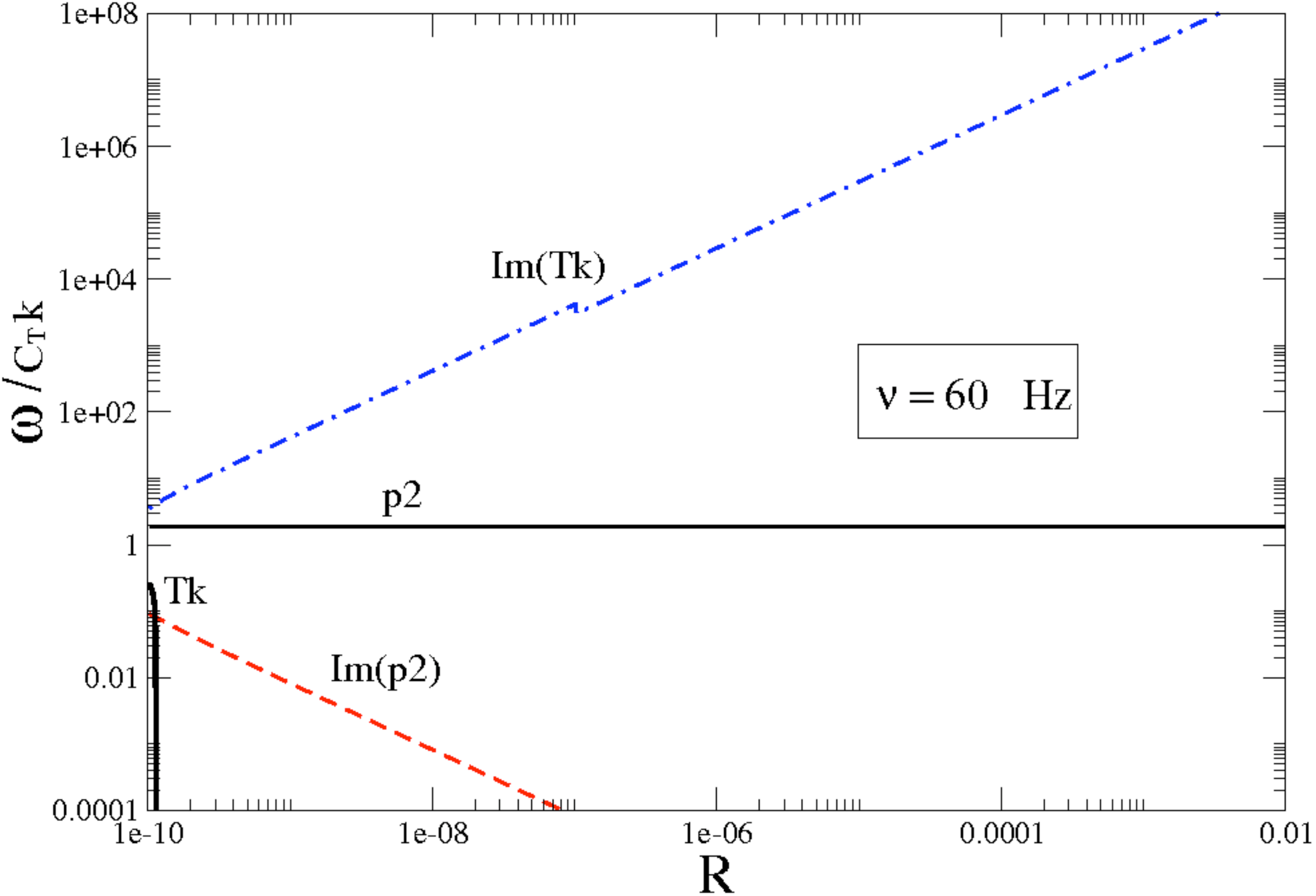}\includegraphics[height=5.5cm,clip]{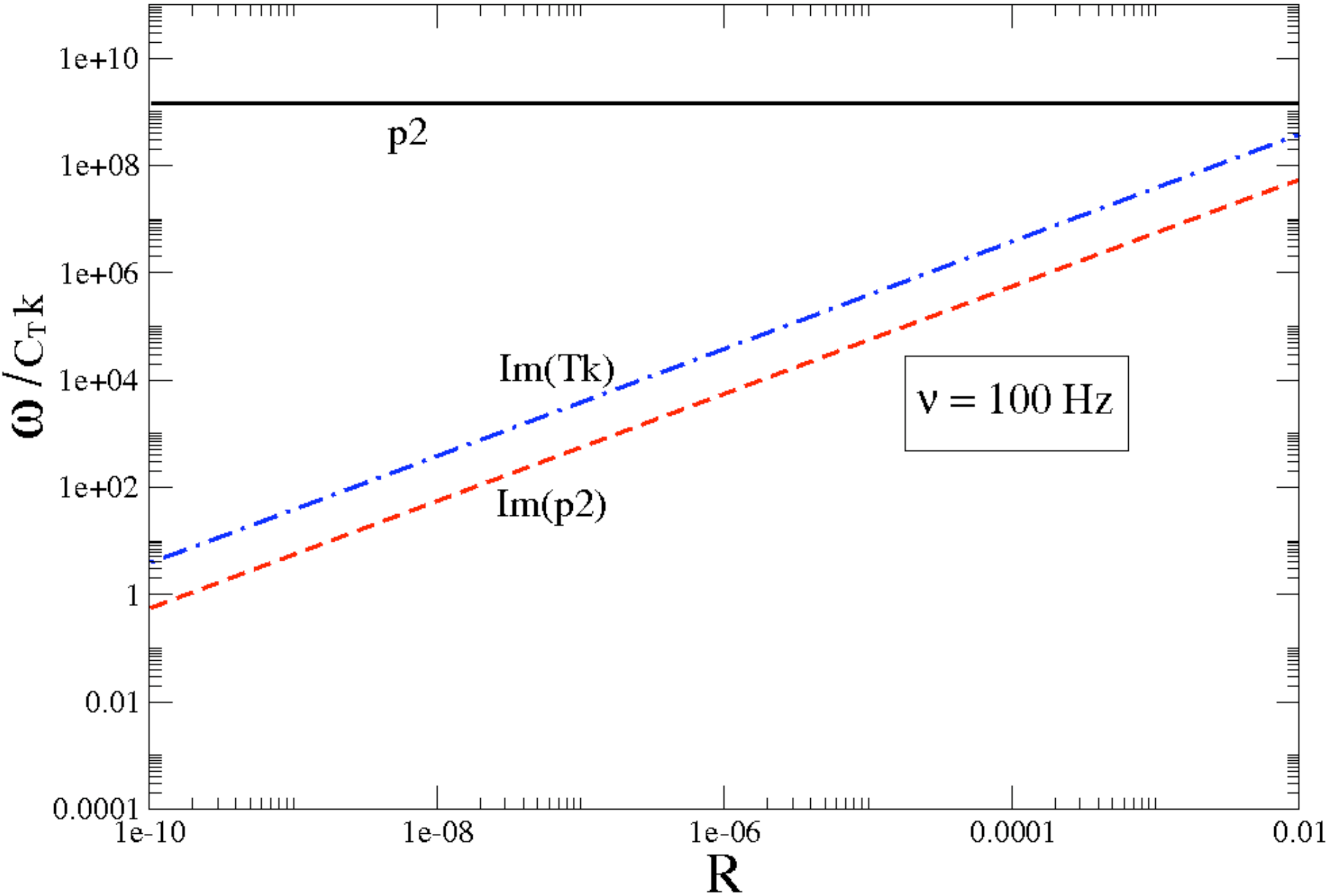}}
\caption{We plot the frequency of the Tkachenko modes and of the sound waves for $\alpha=1.2$ and $\beta=0.8$.  For a rotation rate of 60 Hz one has a sound wave close to the classical Tkachenko wave frequency and a highly damped "soft" Tkachenko wave for low values of $\mathcal{R}$. At 100 Hz, the frequency of the "soft" mode becomes purely imaginary and the sound waves return to being high frequency modes, weakly damped by mutual friction.}
\label{compres2}
\end{figure}

If we now increase the rotation rate in the case $\beta>\alpha$ one finds that the frequency of the Tkachenko waves approaches that of the "soft" mode and mutual friction only weakly damps the mode for realistic values of $x_\p$. The picture is considerably different if we take $\beta<\alpha$, as can be seen from figure (\ref{compres2}). For a rotation rate of 60 Hz one has a sound wave close to the classical Tkachenko wave frequency and a highly damped "soft" Tkachenko wave for low values of $\mathcal{R}$. However, if we increase the rotation rate to 100 Hz, the frequency of the "soft" mode becomes purely imaginary and the sound waves return to being high frequency modes, weakly damped by mutual friction. The picture that emerges is thus that, while for low rotation rates of order of a few Hz one has long lived Tkachenko oscillations for a vast range of plausible parameters (except for the particular case of $\alpha=\beta=1$ as in model A), for higher rotation rates, approaching 100 Hz, the situation is radically different and for several choices of parameters there are no Tkachenko modes at all.

\subsection{Perfect pinning}
\begin{figure}
\centerline{\includegraphics[height=5.5cm,clip]{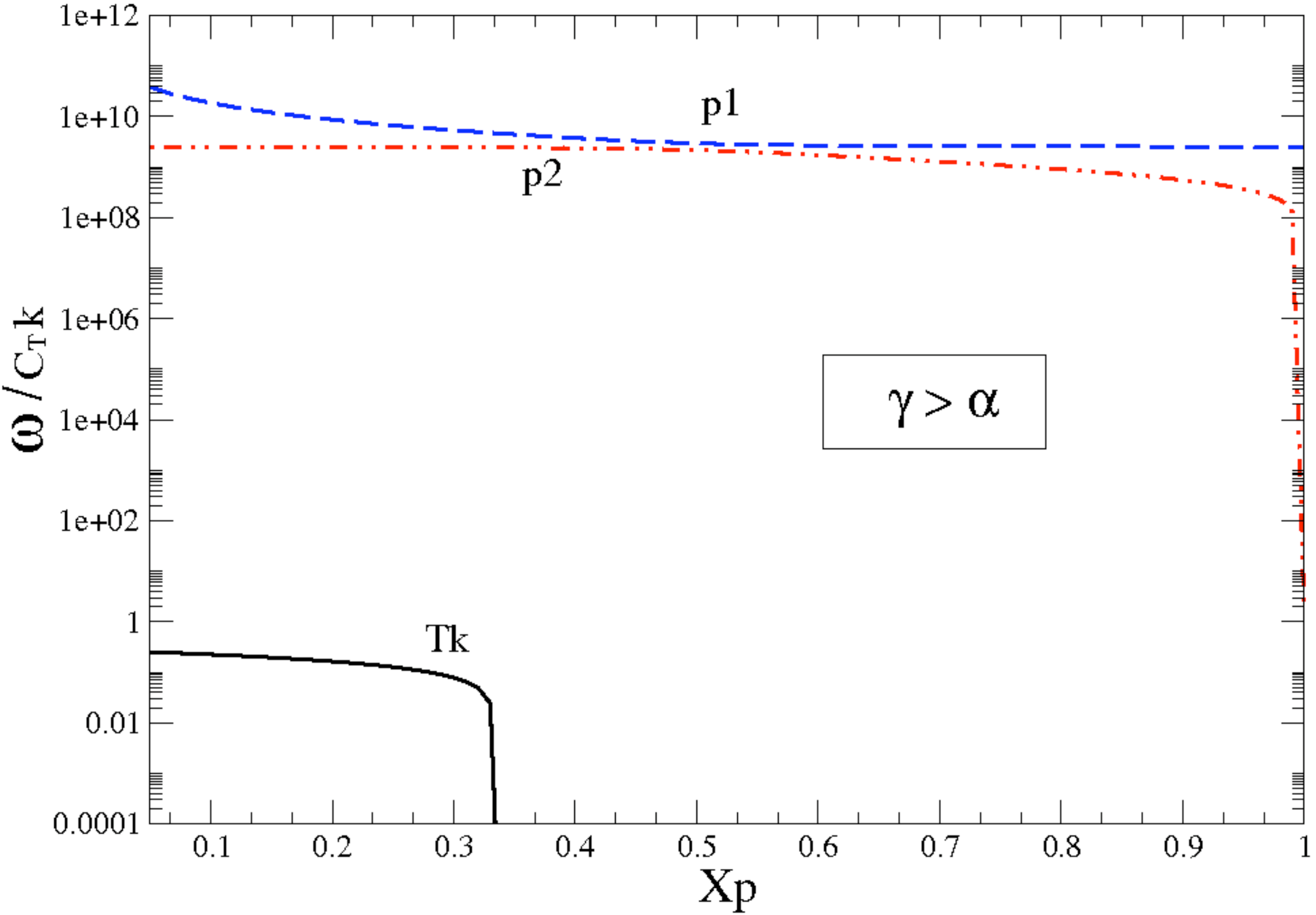}\includegraphics[height=5.5cm,clip]{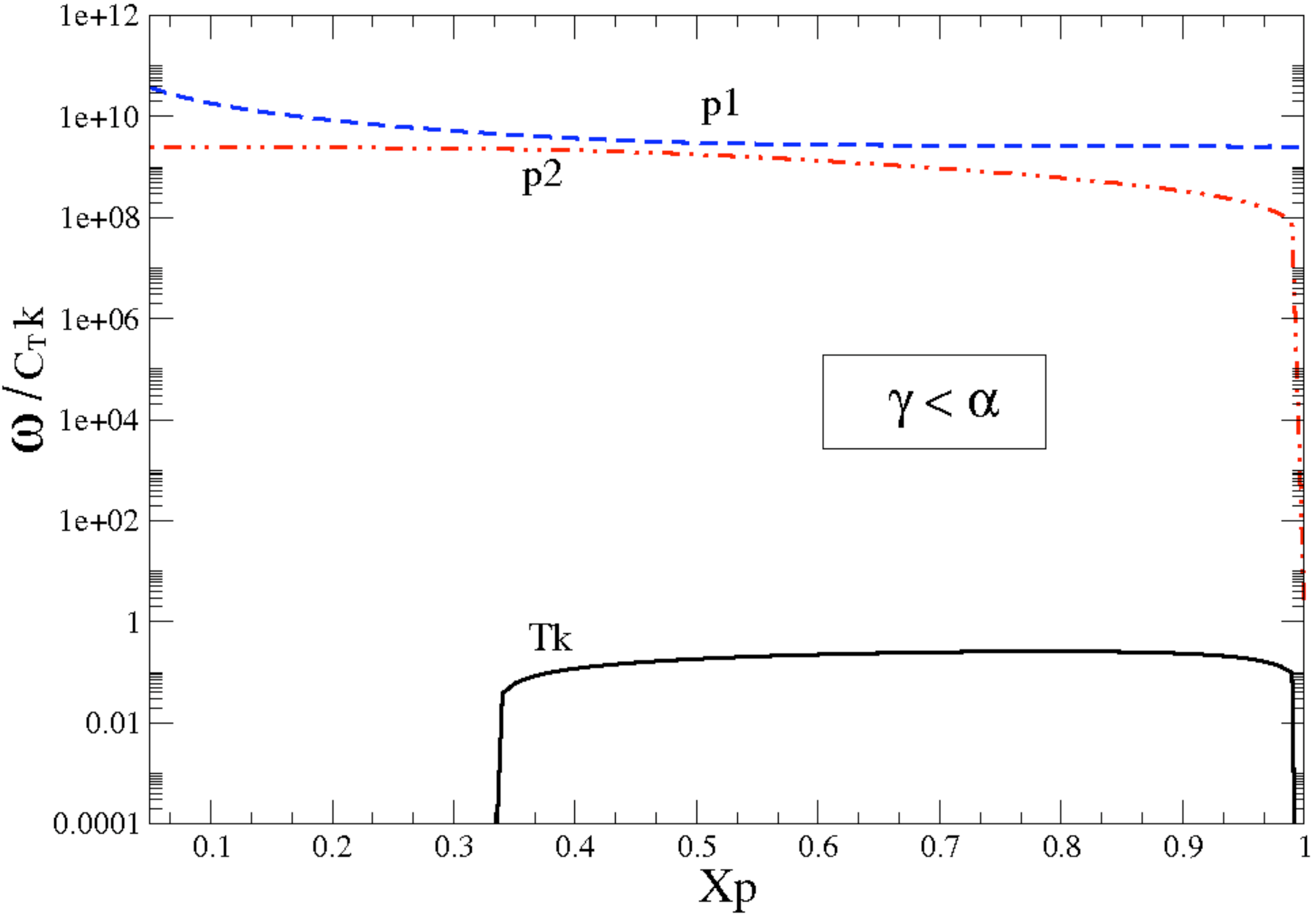}}
\caption{Mode structure for a rotation rate of 100 Hz and varying $x_\p$ in the case of perfect pinning. We consider purely transverse propagation, $k=10^{-6}$ and $\bar{\varepsilon}=0$. For the, possibly more realistic, case of small values of $x_\p$ one has a Tkachenko mode only for $\gamma>\alpha$. However for larger values of $x_\p$ the opposite is true and the Tkachenko mode only exists in the limit $\gamma<\alpha$, but its frequency vanishes for $x_\p\approx 1$.}
\label{pin}
\end{figure}

Finally we examine, as in section III C, the case of perfect pinning. The Euler equations can be written as:
\beq
&&-i\omega\bar{v}_i +i \frac{k_i}{\rho}\bar{p}-\frac{\bar{\rho}}{\rho}\nabla_i p +2\epsilon_{ijk}\Omega^j\bar{v}^k=-(1-x_\p)\sigma_i\\
&&-i(1-\bar{\epsilon})\omega\bar{w}_i+i k_i\bar{\beta}-2\frac{(1-x_\p)}{x_\p}\epsilon_{ijk}\Omega^j\bar{w}^k=-\frac{(1-x_\p)}{x_\p}\sigma_i
\label{Eulkpin}
\eeq
Together with the pinning condition $-i\omega\epsilon_i=\bar{v}_i^\p=\bar{v}_i+(1-x_\p)\bar{w}_i$.
In the limit of no entrainment ($\bar{\varepsilon}=0$)  for purely transverse propagation ($\cos\theta=0$), one finds that, as in the incompressible case, the spectrum depends heavily on the value of the proton fraction $x_\p$. For slow rotation rates (of the order $\approx 10$ Hz) one always has a Tkachenko mode for low values of the proton fraction (less than $x_\p\approx 0.3$). However the situation changes for higher rotation rates. For a stellar rotation rate of 100 Hz, one can see from figure (\ref{pin}) that for the, possibly more realistic, case of small values of $x_\p$ one has a Tkachenko mode only for $\gamma>\alpha$. However for larger values of $x_\p$ the opposite is true and the Tkachenko mode only exists in the limit $\gamma<\alpha$. If $\gamma=\alpha=1$ (model A) one has, as expected, no Tkachenko waves for any value of the proton fraction.

\section{Conclusions}

In this paper we present a formalism for the inclusion of vortex lattice elasticity in the multi-fluid hydrodynamics of superfluid neutron stars. As a first step we consider incompressible neutron and proton matter, in order to make contact with the well known results for superfluid $^4$ He. We obtain the standard dispersion relation for Tkachenko waves and find that, in the limit in which $\rho_\n\approx\rho_\p$, mutual friction completely damps out the oscillations in less than a period when the damping timescale due to the drag is close to the mode period (as found by \cite{Sed1}). However in a neutron star one will, in general, have that $\rho_\p << \rho_\n$, and in this case we always find weakly damped oscillatory solutions.
It should be stressed that a more realistic model should include other sources of damping, such as shear and bulk viscosity, and the calculation should be performed for global modes in spherical symmetry, without assuming constant background quantities.

The main focus of this work is on the effect of compressibility and chemical coupling on the mode spectrum. We find that for slow rotation rates (a few Hz) one can, in general, obtain solutions that correspond to low frequency Tkachenko waves and could explain the timing noise in older pulsars. However, for particular choices of the EOS parameters, such as model A, there are no Tkachenko waves, but only modified inertial waves and sound waves, that oscillate at frequencies that are to high to have any connection with the timing noise. Finally the situation is considerably more complicated for more rapidly roatating NSs (above 100 Hz) for which there is now a large portion of parameter space for which there are no propagating Tkachenko waves. 
It is thus clearly imperative to obtain more stringent constraints on the EOS from nuclear physics, in order to understand if the regions of parameter space in which one has no Tkachenko waves are of physical significance or not. 
In the light of these uncertainties, it is still very much an open question whether or not the period of Tkachenko waves in a realistic neutron star could explain the observed periodicity of $\approx 256$ days in the timing residuals of PSR B1828-11 (which is rotating at $\nu=2.469$ Hz) and the timing noise in other pulsars, or if it is likely to power low frequency precessional motion of the star. 
Our results, however, indicate that for a large range of physical parameters long period Tkachenko waves can in fact propagate in a NS interior and are likely to play a role in the dynamics of the system.

Finally let us remark that we have considered perturbations of a co-moving background. While this is not a bad approximation in many situations, it is possible that if the vortices are pinned to the crust a significant lag could build up between the charged component and the superfluid neutrons. This can lead to a series of short wave-length instabilities (\cite{NT},\cite{NK}) that are likely to have an impact on pulsar glitches and on the stellar response to external torques, such as those experienced by neutron stars in accreting systems. We plan to relax the assumption of a co-moving background and explore the consequences on these physical scenarios in future work.

\section*{Acknowledgments}

I wish to thank N.Andersson and D.I.Jones for useful discussions.
I acknowledge support from STFC via grant number PP/EE001025/1 and via a EU Marie-Curie Intra-European Fellowship, project number 252470, AMXP dynamics.
This work was partially supported by CompStar, a Research Training Network Programme of the European Science Foundation

\appendix
\section{Characteristic equation: the incompressible case}

For the incompressible case the equations of motion (\ref{eulersk1}-\ref{eulersk}) can be cast in the form:
\be
K_{ij}=\left(\begin{array}{c c c c c c c c c c}
 & \mathcal{A}^{\n}_{ij} & & & &\mathcal{C}^{\n\p}_{ij}& & & \mathcal{E}^{\n}_{ij} &\\
 & & & & & & & & &\\
  & \mathcal{C}^{\p\n}_{ij} & & & &\mathcal{A}^{\p}_{ij}& & & \mathcal{E}^{\p}_{ij} &\\
   & & & & & & & & &\\
  & \mathcal{W}^{\n}_{ij}&  & & &\mathcal{W}^{\p}_{ij} & & & \mathcal{T}_{ij}  
 \end{array}\right)\left(\begin{array}{c}\mathcal{V}^j_\n\\  \\ \mathcal{V}^j_\p \\  \\ \epsilon^j\end{array}\right)\label{kappa1}
 \ee
 
 where
 \be
 \mathcal{V}^j_\n=(\bar{v}^j_\n , \bar{\mu}_\n)\;\;\;\;\mbox{and}\;\;\;\;  \mathcal{V}^j_\p=(\bar{v}^j_\p , \bar{\mu}_\p)
\ee 
 and we recall that the displacement $\epsilon^j$ only has components in the plane perpendicular to the vortices, i.e. the $x-y$ plane in our formulation.
 The components of the matrix $K_{ij}$ take the explicit form:
 \be
 \mathcal{A}^\n_{ij}=\left(\begin{array}{c c c c c c c}
 -i\omega (1-\varepsilon_\n) & &-2\Omega& & 0 & & i k\sin\theta \\
  & & & & & & \\
 2\Omega & & -i\omega (1-\varepsilon_\n) & & 0 & & 0\\
  & & & & & & \\
 0 & & 0 & & -i\omega(1-\varepsilon_\n) & & i k\cos\theta\\
 & & & & & & \\
 k\sin\theta & & 0 & & k\cos\theta & & 0
 \end{array}\right),
 \ee
 \be
 \mathcal{A}^\p_{ij}=\left(\begin{array}{c c c c c c c}
 -i\omega (1-\varepsilon_\p)+2\mathcal{R}\Omega\left(\frac{1-x_\p}{x_\p}\right) & &-2\Omega& & 0 & & i k\sin\theta \\
  & & & & & & \\
 2\Omega & & -i\omega (1-\varepsilon_\p)+2\mathcal{R}\Omega\left(\frac{1-x_\p}{x_\p}\right)  & & 0 & & 0\\
  & & & & & & \\
 0 & & 0 & & -i\omega(1-\varepsilon_\p)+2\mathcal{R}\Omega\left(\frac{1-x_\p}{x_\p}\right)  & & i k\cos\theta\\
 & & & & & & \\
 k\sin\theta & & 0 & & k\cos\theta & & 0
 \end{array}\right),
 \ee
 \be
 \mathcal{C}^{\n\p}_{ij}=\left(\begin{array}{c c c c c c c}
 -i\omega\varepsilon_\n-2\mathcal{R}\Omega& & 0 & & 0 & &0 \\
  & & & & & & \\
0 & &  -i\omega\varepsilon_\n-2\mathcal{R}\Omega& & 0 & & 0\\
  & & & & & & \\
 0 & & 0 & &  -i\omega\varepsilon_\n-2\mathcal{R}\Omega & & 0\\
 & & & & & & \\
 0 & & 0 & &0 & & 0
 \end{array}\right),
 \ee
 \be
 \mathcal{C}^{\p\n}_{ij}=\left(\begin{array}{c c c c c c c}
 -i\omega\varepsilon_\p& & 0 & & 0 & &0 \\
  & & & & & & \\
0 & &  -i\omega\varepsilon_\p& & 0 & & 0\\
  & & & & & & \\
 0 & & 0 & &  -i\omega\varepsilon_\p & & 0\\
 & & & & & & \\
 0 & & 0 & &0 & & 0
 \end{array}\right),
 \ee

\be
 \mathcal{W}^{\n}_{ij}=\left(\begin{array}{c c c c c }
 -\frac{1}{1+\mathcal{R}^2}& & -\mathcal{B} & & 0\\
  & & &  &\\
\mathcal{B} & &   -\frac{1}{1+\mathcal{R}^2}& & 0
  \end{array}\right),
\ee

\be
 \mathcal{W}^{\p}_{ij}=\left(\begin{array}{c c c c c }
 -\frac{\mathcal{R}^2}{1+\mathcal{R}^2}& & \mathcal{B} & & 0\\
  & & &  &\\
-\mathcal{B} & &   -\frac{\mathcal{R}^2}{1+\mathcal{R}^2}& & 0
  \end{array}\right),
\ee

\be
 \mathcal{E}^{\n}_{ij}=\left(\begin{array}{c c c }
 -c_T^2(k\sin\theta)^2-2i\omega\mathcal{R}\Omega & & 0\\
  & & \\
0 & & c_T^2(k\sin\theta)^2-2i\omega\mathcal{R}\Omega 
  \end{array}\right),
\ee

\be
 \mathcal{E}^{\p}_{ij}=\left(\begin{array}{c c c }
 2i\omega\mathcal{R}\frac{(1-x_\p)}{x_\p}\Omega & & 0\\
  & & \\
0 & &2i\omega\mathcal{R}\frac{(1-x_\p)}{x_\p}\Omega 
  \end{array}\right),
\ee

\be
 \mathcal{T}_{ij}=\left(\begin{array}{c c c }
 -i\omega-\frac{c_T^2(k\sin\theta)^2\mathcal{B}}{2\Omega} & &-\frac{c_T^2(k\sin\theta)^2}{2\Omega(1+\mathcal{R}^2)}  \\
  & & \\
-\frac{c_T^2(k\sin\theta)^2}{2\Omega(1+\mathcal{R}^2)} & & -i\omega+\frac{c_T^2(k\sin\theta)^2\mathcal{B}}{2\Omega} 
  \end{array}\right).
\ee
where we recall the definition $\mathcal{B}=\mathcal{R}/(1+\mathcal{R}^2)$. The characteristic equation then follows from the determinant of the matrix $K_{ij}$.

\subsection{perfect pinning}\label{p1}

In the case of perfect pinning the elements of the matrix $K_{ij}$ are modified in the following way:

 \be
 \mathcal{A}^\n_{ij}=\left(\begin{array}{c c c c c c c}
 -i\omega (1-\varepsilon_\n) & & 0 & & 0 & & i k\sin\theta \\
  & & & & & & \\
 0 & & -i\omega (1-\varepsilon_\n) & & 0 & & 0\\
  & & & & & & \\
 0 & & 0 & & -i\omega(1-\varepsilon_\n) & & i k\cos\theta\\
 & & & & & & \\
 k\sin\theta & & 0 & & k\cos\theta & & 0
 \end{array}\right),\label{ap2}
 \ee
 \be
 \mathcal{A}^\p_{ij}=\left(\begin{array}{c c c c c c c}
 -i\omega (1-\varepsilon_\p) & &2\Omega\left(\frac{1-2 x_\p}{x_\p}\right)& & 0 & & i k\sin\theta \\
  & & & & & & \\
 -2\Omega\left(\frac{1-2x_\p}{x_\p}\right) & & -i\omega (1-\varepsilon_\p)  & & 0 & & 0\\
  & & & & & & \\
 0 & & 0 & & -i\omega(1-\varepsilon_\p) & & i k\cos\theta \\
 & & & & & & \\
 k\sin\theta & & 0 & & k\cos\theta & & 0
 \end{array}\right),
 \ee
 \be
 \mathcal{C}^{\n\p}_{ij}=\left(\begin{array}{c c c c c c c}
 -i\omega\varepsilon_\n& & -2\Omega & & 0 & &0 \\
  & & & & & & \\
2\omega & &  -i\omega\varepsilon_\n& & 0 & & 0\\
  & & & & & & \\
 0 & & 0 & &  -i\omega\varepsilon_\n & & 0\\
 & & & & & & \\
 0 & & 0 & &0 & & 0
 \end{array}\right),
 \ee
 \be
 \mathcal{C}^{\p\n}_{ij}=\left(\begin{array}{c c c c c c c}
 -i\omega\varepsilon_\p& & -2\Omega\left(\frac{1-x_\p}{x_\p}\right) & & 0 & &0 \\
  & & & & & & \\
2\Omega\left(\frac{1-x_\p}{x_\p}\right) & &  -i\omega\varepsilon_\p& & 0 & & 0\\
  & & & & & & \\
 0 & & 0 & &  -i\omega\varepsilon_\p & & 0\\
 & & & & & & \\
 0 & & 0 & &0 & & 0
 \end{array}\right),
 \ee

\be
 \mathcal{W}^{\p}_{ij}=\left(\begin{array}{c c c c c }
1& & 0 & & 0\\
  & & &  &\\
0 & &   1 & & 0
  \end{array}\right),\;\;\;\;\mbox{}\;\;\;\;  \mathcal{T}_{ij}=\left(\begin{array}{c c c }
 i\omega& &0  \\
  & & \\
0 & & i\omega
  \end{array}\right).
\ee

\be
 \mathcal{E}^{\n}_{ij}=0 \;\;\;\;\mbox{}\;\;\;\; \mathcal{W}^{\n}_{ij}=0,
 \ee

\be
 \mathcal{E}^{\p}_{ij}=\left(\begin{array}{c c c }
-\frac{(1-x_\p)}{x_\p}c_T^2(k\sin\theta)^2 & & 0\\
  & & \\
0 & & \frac{(1-x_\p)}{x_\p}c_T^2(k\sin\theta)^2 
  \end{array}\right),\label{ap22}
\ee

\section{Characteristic equation: the compressible case}\label{comprimo}

In the compressible case we need to cast the equations of motion  (\ref{eulersk1}-\ref{eulersk}) in the form:
\be
K_{ij}=\left(\begin{array}{c c c c c c c c c c}
 & \mathcal{A}^{T}_{ij} & & & &\mathcal{C}^{TC}_{ij}& & & \mathcal{E}^{T}_{ij} &\\
 & & & & & & & & &\\
  & \mathcal{C}^{CT}_{ij} & & & &\mathcal{A}^{C}_{ij}& & & \mathcal{E}^{C}_{ij} &\\
   & & & & & & & & &\\
  & \mathcal{W}^{T}_{ij}&  & & &\mathcal{W}^{C}_{ij} & & & \mathcal{T}_{ij}  
 \end{array}\right)\left(\begin{array}{c}\mathcal{V}^j_T\\  \\ \mathcal{V}^j_C \\  \\ \epsilon^j\end{array}\right)\label{ap3}
 \ee
 
 where
 \be
 \mathcal{V}^j_T=(\bar{v}^j_T , \bar{\rho}_T)\;\;\;\;\mbox{and}\;\;\;\;  \mathcal{V}^j_C=(\bar{v}^j_C , \bar{x}_\p)
\ee 
 The components of the matrix $K_{ij}$ take the explicit form:
 \be
 \mathcal{A}^T_{ij}=\left(\begin{array}{c c c c c c c}
 -i\omega & &-2\Omega& & 0 & & i k\sin\theta\frac{c_s^2}{\rho} \\
  & & & & & & \\
 2\Omega & & -i\omega & & 0 & & 0\\
  & & & & & & \\
 0 & & 0 & & -i\omega & & i k\cos\theta\frac{c_s^2}{\rho}\\
 & & & & & & \\
 i \rho k\sin\theta & & 0 & &i \rho k\cos\theta & & -i\omega
 \end{array}\right),
 \ee
 \be
 \mathcal{A}^C_{ij}=\left(\begin{array}{c c c c c c c}
 -i\omega (1-\bar{\varepsilon})+2\mathcal{R}\Omega\left(\frac{1-x_\p}{x_\p}\right) & &-2\Omega& & 0 & & i k\sin\theta\frac{c_s^2}{x_\p^2} \\
  & & & & & & \\
 2\Omega & & -i\omega (1-\bar{\varepsilon})+2\mathcal{R}\Omega\left(\frac{1-x_\p}{x_\p}\right)  & & 0 & & 0\\
  & & & & & & \\
 0 & & 0 & & -i\omega(1-\bar{\varepsilon})+2\mathcal{R}\Omega\left(\frac{1-x_\p}{x_\p}\right)  & & i k\cos\theta\frac{c_s^2}{x_\p^2}\\
 & & & & & & \\
 i x_\p(1-x_\p)k\sin\theta & & 0 & &  i x_\p(1-x_\p)k\cos\theta & & -i\omega
 \end{array}\right),
 \ee
 \be
 \mathcal{C}^{TC}_{ij}=\left(\begin{array}{c c c c c c c}
0& & 0 & & 0 & &i k\sin\theta\frac{c_s^2}{x_\p} \\
  & & & & & & \\
0 & & 0& & 0 & & 0\\
  & & & & & & \\
 0 & & 0 & &  0 & & i k\cos\theta\frac{c_s^2}{x_\p}\\
 & & & & & & \\
 0 & & 0 & &0 & & 0
 \end{array}\right),
 \ee
 \be
 \mathcal{C}^{CT}_{ij}=\left(\begin{array}{c c c c c c c}
2\Omega\frac{\mathcal{R}}{x_\p}& & 0 & & 0 & &i k\sin\theta\frac{c_s^2}{\rho x_\p}\\
  & & & & & & \\
0 & &  2\Omega\frac{\mathcal{R}}{x_\p}& & 0 & & 0\\
  & & & & & & \\
 0 & & 0 & & 2\Omega\frac{\mathcal{R}}{x_\p}& & i k\cos\theta\frac{c_s^2}{\rho x_\p}\\
 & & & & & & \\
 0 & & 0 & &0 & & 0
 \end{array}\right),
 \ee

\be
 \mathcal{W}^{T}_{ij}=\left(\begin{array}{c c c c c }
 -1& & 0& & 0\\
  & & &  &\\
0 & &   -1& & 0
  \end{array}\right),
\ee

\be
 \mathcal{W}^{C}_{ij}=\left(\begin{array}{c c c c c }
 -(1-x_\p)+\frac{1}{1+\mathcal{R}^2}& & \mathcal{B} & & 0\\
  & & &  &\\
-\mathcal{B} & &   -(1-x_\p)+\frac{1}{1+\mathcal{R}^2}& & 0
  \end{array}\right),
\ee

\be
 \mathcal{E}^{T}_{ij}=\left(\begin{array}{c c c }
 -(1-x_\p)c_T^2(k\sin\theta)^2 & & 0\\
  & & \\
0 & & (1-x_\p)c_T^2(k\sin\theta)^2
  \end{array}\right),
\ee

\be
 \mathcal{E}^{C}_{ij}=\left(\begin{array}{c c c }
 2i\omega\frac{\mathcal{R}}{x_\p}\Omega+\frac{(1-x_\p)}{x_\p}c_T^2(k\sin\theta)^2 & & 0\\
  & & \\
0 & &2i\omega\frac{\mathcal{R}}{x_\p}\Omega- \frac{(1-x_\p)}{x_\p}c_T^2(k\sin\theta)^2
  \end{array}\right),
\ee

\be
 \mathcal{T}_{ij}=\left(\begin{array}{c c c }
 -i\omega-\frac{c_T^2(k\sin\theta)^2\mathcal{B}}{2\Omega} & &-\frac{c_T^2(k\sin\theta)^2}{2\Omega(1+\mathcal{R}^2)}  \\
  & & \\
-\frac{c_T^2(k\sin\theta)^2}{2\Omega(1+\mathcal{R}^2)} & & -i\omega+\frac{c_T^2(k\sin\theta)^2\mathcal{B}}{2\Omega} 
  \end{array}\right).
\ee

\subsection{perfect pinning}

In the case of perfect pinning, for a compressible model, the elements of the matrix $K_{ij}$ are modified in the following way:

 \be
 \mathcal{A}^T_{ij}=\left(\begin{array}{c c c c c c c}
 -i\omega & & -2\Omega & & 0 & & i k\sin\theta\frac{c_s^2}{\rho} \\
  & & & & & & \\
 2\Omega & & -i\omega & & 0 & & 0\\
  & & & & & & \\
 0 & & 0 & & -i\omega& & i k\cos\theta\frac{c_s^2}{\rho}\\
 & & & & & & \\
 i\rho k\sin\theta & & 0 & & i\rho k\cos\theta & & -i\omega
 \end{array}\right),
 \ee
 
 \be
 \mathcal{A}^C_{ij}=\left(\begin{array}{c c c c c c c}
 -i\omega (1+\bar{\varepsilon}) & &2\Omega\left(\frac{1- x_\p}{x_\p}\right)& & 0 & & i k\sin\theta\frac{c_s^2}{x_\p^2} \\
  & & & & & & \\
 -2\Omega\left(\frac{1-x_\p}{x_\p}\right) & & -i\omega (1+\bar{\varepsilon})  & & 0 & & 0\\
  & & & & & & \\
 0 & & 0 & & -i\omega(1+\bar{\varepsilon}) & & i k\cos\theta\frac{c_s^2}{x_\p^2}\\
 & & & & & & \\
 i x_\p(1-x_\p) k\sin\theta & & 0 & & i x_\p (1-x_\p)k\cos\theta & & -i\omega
 \end{array}\right),
\ee

\be
 \mathcal{C}^{TC}_{ij}=\left(\begin{array}{c c c c c c c}
0& & 0 & & 0 & &i k\sin\theta\frac{c_s^2}{x_\p} \\
  & & & & & & \\
0 & & 0& & 0 & & 0\\
  & & & & & & \\
 0 & & 0 & &  0 & & i k\cos\theta\frac{c_s^2}{x_\p}\\
 & & & & & & \\
 0 & & 0 & &0 & & 0
 \end{array}\right),
 \ee
 
 \be
 \mathcal{C}^{CT}_{ij}=\left(\begin{array}{c c c c c c c}
0& & 0 & & 0 & &i k\sin\theta\frac{c_s^2}{\rho x_\p}\\
  & & & & & & \\
0 & &  0& & 0 & & 0\\
  & & & & & & \\
 0 & & 0 & & 0& & i k\cos\theta\frac{c_s^2}{\rho x_\p}\\
 & & & & & & \\
 0 & & 0 & &0 & & 0
 \end{array}\right),
 \ee

\be
 \mathcal{W}^{T}_{ij}=\left(\begin{array}{c c c c c }
 -1& & 0& & 0\\
  & & &  &\\
0 & &   -1& & 0
  \end{array}\right),
\;\;\;\;\mbox{}\;\;\;\;
 \mathcal{W}^{C}_{ij}=\left(\begin{array}{c c c c c }
 -(1-x_\p)& & 0 & & 0\\
  & & &  &\\
0 & &   -(1-x_\p)& & 0
  \end{array}\right),
\ee

\be
 \mathcal{E}^{T}_{ij}=\left(\begin{array}{c c c }
 -(1-x_\p)c_T^2(k\sin\theta)^2 & & 0\\
  & & \\
0 & & (1-x_\p)c_T^2(k\sin\theta)^2
  \end{array}\right),
\ee

\be
 \mathcal{E}^{C}_{ij}=\left(\begin{array}{c c c }
 -\frac{(1-x_\p)}{x_\p}c_T^2(k\sin\theta)^2 & & 0\\
  & & \\
0 & &\frac{(1-x_\p)}{x_\p}c_T^2(k\sin\theta)^2
  \end{array}\right),
\ee

\be
\mathcal{T}_{ij}=\left(\begin{array}{c c c }
- i\omega& &0  \\
  & & \\
0 & & -i\omega
  \end{array}\right).
\ee

\end{document}